\documentstyle[12pt,aasms,tighten]{article}
\begin{document}

\title{The Dynamics and Outcomes of Rapid Infall onto Neutron Stars}
\author{Chris L. Fryer, Willy Benz}

\affil{Steward Observatory, University of Arizona, Tucson, AZ 85721
e-mail:cfryer@as.arizona.edu}
\and
\author{Marc Herant}
\affil{Theory Division MS K710, Los Alamos National Laboratory,
Los Alamos, NM 87545}

\begin{abstract}

We present an extensive study of accretion onto neutron stars in which
the velocity of the neutron star and structure of the surrounding
medium is such that the Bondi-Hoyle accretion exceeds $10^{-3} M_{\odot}$
y$^{-1}$. Two types of initial conditions are considered for a range
of entropies and chemical compositions: an atmosphere in pressure
equilibrium above the neutron star, and a freely falling inflow of
matter from infinity (also parametrized by the infall rate). We then
evolve the system with one- and two-dimensional hydrodynamic codes
to determine the outcome.  For most cases, hypercritical (also termed ``super
Eddington'') accretion due to rapid neutrino cooling allows the neutron
star to accrete above the Bondi-Hoyle rate as previously pointed out
by Chevalier.  However, for a subset of simulations which corresponds to
evolutionarily common events, convection driven by neutrino heating
can lead to explosions by a mechanism similar to that found in
core-collapse supernovae.

Armed with the results from our calculations, we are in a position
to predict the fate of a range of rapid-infall neutron star
accretors present in certain low-mass X-ray binaries, common envelope
systems, supernova fallbacks and Thorne-Zytkow objects (TZOs).  A
majority of the common envelope systems that we considered led to
explosions expelling the envelope, halting the neutron star's inward
spiral, and allowing the formation of close binary systems.  As a
result, the smothered neutron stars produced in the collisions
studied by Davies \& Benz may also explode, probably preventing
them from forming millisecond pulsars. For the most massive supernovae,
in which the fallback of material towards the neutron star after a
successful explosion is large, we find that a black hole is
formed in a few seconds.  Finally, we argue that the current set
of TZO formation scenarios is inadequate and leads instead to
hypercritical accretion and black hole formation.  Moreover, it
appears that many of the current TZ models have structures
ill-suited for modeling by mixing length convection. This has
prompted us to develop a simple test to determine the viability
of this approximation for a variety of convective systems.

\end{abstract}


\section{INTRODUCTION} \label{sec:int}

It is only in the last few decades, with the arrival of high-energy
observatories, that the problem of accretion onto neutron stars has
moved from the speculations of theorists to the constraints of
observations.  Satellites such as {\it Einstein}, {\it ROSAT},
{\it GRO}, {\it Ginga}, and others, have contributed to a growing
list of accreting neutron star sources such as gamma-ray bursters,
X-ray bursters, millisecond pulsars, high mass X-ray binaries (HMXB),
low mass X-ray binaries (LMXB), and a number of objects entangled
within the current evolutionary scenarios for binary pulsars.
Future observations ({\it AXAF}, {\it NAE}, etc.) promise
to add more.  Unfortunately, the current state of theoretical models
falls short of the present and upcoming data.  At the root of the
theoretical difficulties is the range of extreme physical conditions
encountered in many of the observed systems:  high magnetic fields,
angular momentum, degenerate matter, neutrino effects, etc.  In
addition, as we shall demonstrate in this paper, it is likely that
multidimensional effects are important.  As a result,
progress in understanding neutron star accretion has been
slow.  In this paper, the first of a series, we will
consider the effects of rapid mass infall onto neutron stars
($ \dot{M}_{\rm Bondi-Hoyle}>10^{-3}M_{\odot}$ y$^{-1}$).  By {\it infall}
rate, we mean the rate at which material is added to the atmosphere
surrounding the neutron star. This is to be distinguished from the term
{\it accretion} which we reserve for the mass incorporated into
the neutron star.  High infall rates occur in common envelope systems such
as Be/X-ray objects, more deeply buried systems such as Thorne-Zytkow
objects (TZOs), and supernova fallback.

Early work studying rapid mass infall onto neutron stars logically
began with estimates of the fallback of matter onto newborn
neutron stars in supernovae (Colgate 1971; Zel'dovich, Ivanova,
\& Nadezhin 1972).  Both groups found that the canonical photon
Eddington accretion rate vastly underestimates accretion onto
the neutron star as neutrino, rather than photon, emission becomes
the dominant cooling source.  Chevalier (1989), and Houck \&
Chevalier (1991) have studied in greater analytic
detail the fallback of matter onto the adolescent neutron star,
confirming this ``hypercritical'' accretion rate.  Even though
the amount of fallback matter is generally a small fraction of
the material expelled by the supernova, it is a large portion of
the material undergoing heavy element nucleosynthesis.  Thus,
understanding supernova fallback is crucial, not only to decide
whether or not a neutron star or black hole is left after the
explosion, but also to understand the nucleosynthetic yields of
supernova explosions which, in turn, have profound repercussions
for galactic chemical evolution.

However, most of the interest in atmospheres around neutron stars has
been directed toward understanding Thorne-Zytkow objects, an hypothetical
red giant in which the normally white dwarf-like core is a neutron star.
These objects are supported, in addition to the ``normal'' contribution
from thermonuclear burning, by the release of gravitational energy from
accretion onto the neutron star, and therefore have longer lifetimes
than standard red giants.  The concept of powering a star from mass
accretion onto a degenerate object was revived from its pre-fusion days
(Landau 1937; Gamow 1937) by Thorne \& Zytkow (1975).  In their study of
a range of models for stellar envelopes greater than 1 $M_\odot$, Thorne
\& Zytkow (1977) separated their models into two classes depending upon
atmosphere mass.  For the more massive envelopes, accretion alone is
insufficient to provide pressure support, suggesting nuclear burning
arising from the extreme conditions near the surface of the neutron
star as a possible complementary mechanism.  Follow-up work by
Biehle (1991, 1994), Cannon et al. (1992), and Cannon (1993) focused
on this class of objects, using more detailed descriptions of the
nuclear burning.  These simulations established abnormal nuclear burning
(such as the rp-process) as
an additional source of pressure support and led to more definitive
results on the observable chemical compositions of TZOs, should they
exist.  In recent years, the usage of the terms TZO has grown
in the literature to encompass a broader range of neutron-star
atmosphere systems.  In this paper, we reserve the TZO designation
for those objects originally envisioned by Thorne and Zytkow and
the other authors listed in this paragraph.

As an ever-branching list of formation scenarios has been dreamed up,
the occurence of rapid infall on neutron stars has evolved from the seed
of a theorist's imagination to a virtual certainty. Among them one
finds: 1) supernova fallback, 2) common envelope evolution (Taam,
Bodenheimer, \& Ostriker 1978; Terman, Taam, \& Hernquist 1994),
3) collisions between main-sequence stars or red giants and neutron
stars in globular clusters or galactic nuclei (Benz \& Hills 1992;
Davies \& Benz 1995), 4) an induced collision between a newly formed
neutron star and its binary companion as it is kicked by an asymmetric
supernova explosion (Leonard, Hills, \& Dewey 1994), and 5) a neutron
star caught within the torus of an active galactic nucleus or within
a dense molecular cloud (D. N. C. Lin - private communication).
Supernova fallback excepted, not much attention has been paid
to the link between formation scenario and their hypothetical product
(LMXB, TZO, etc.). Along those lines, the stability of the build-up
of an atmosphere around a neutron star has been in dispute ever since
TZ objects were conjectured.  At the heart of this controversy is the
impact of neutrino physics on the structure of the atmosphere.
Neutrino cooling, an aspect of the problem first brought up by
Bisnovatyi-Kogan \& Lamzin (1984), dismissed by Eich et
al. (1989) and subsequently ignored by Biehle (1991, 1994),
Cannon et al. (1992) and Cannon (1993) has regained prominence
as understanding of supernova fallback onto neutron stars has
progressed. In particular, Chevalier (1993) has found that
hypercritical accretion can occur, not only in the case of
supernova fallback, but also in the entire range of common
envelope systems, including TZ objects.

This paper is about what occurs when a neutron star is forced to
accrete matter at a high rate.  A generic scenario might be as
follows: a neutron star encounters a star (or a gas cloud, or an
AGN disk, or its own ejecta). During an initial transient, an
accretion shock moves from the surface of the neutron star to
some equilibrium radius. The region inside the shock then becomes
an atmosphere in near-pressure equilibrium which settles on the
neutron star. Both of these phases of the evolution can last for
many dynamical times so that it is not possible to model the
comprehensive evolution of the accreting neutron star. Since we
would still like to determine what will be the final outcome, we
have devised the following plan of attack: in a first set of
simulations, we compute the evolution of a series of initial
free-fall conditions over a range of parameters (entropy, chemical
composition, and most importantly, infall rate).  These conditions
typify the expected initial structure obtained as a neutron star
plows into a medium. For reasons explained in \S 4.1.1, the
transient structure which develops while the accretion shock moves
outward immediately becomes convectively unstable, pushing the
accretion shock beyond its steady state radius.
Unfortunately, the convective episode lasts too long to follow to
completion, but the end result can be inferred, namely that an
isentropic atmosphere will eventually build up above the neutron star.
In a second set of simulations, we examine just such isentropic
atmospheres initially in gravitational and pressure equilibrium
for a range of entropy and chemical composition.  In the absence of
neutrino processes, these atmospheres are stable. Starting with these
``pseudo-stable'' initial conditions, we turn on neutrino processes
and determine how stability is affected by energy losses due to
neutrino emission.  Combining these two sets of simulations, we
can take a given infall structure and, using our first set of
models, estimate the resulting structure of the isentropic atmosphere.
With this structure, the second series of models will then predict
the ultimate fate of the system.

Section 2 discusses the numerical methods used in our simulations
and \S 3 presents the
range of initial conditions and physical processes included in this
study.  The results, including comparisons to the work of Chevalier
(1989, hereafter C89), Chevalier (1993), Houck \& Chevalier (1991)
(hereafter HC), and Colgate, Herant, \& Benz (1993) (hereafter CHB), are
presented in \S 4.  We conclude with a discussion of the
implications of our findings for the variety of systems involving
high mass infall and the potential observational consequences.

\section{NUMERICAL METHODS}

To explore the behavior of our range of atmospheres, we utilize both
one-dimensional and two-dimensional codes.  The one-dimensional code
is used to determine the subset of atmospheres that develop negative
entropy gradients of sufficient magnitudes to induce convection and
thus break the otherwise spherical symmetry.  This allows us to limit
two-dimensional simulations to those atmospheres that really require
it.  All physical processes such as neutrino physics (absorption,
emission and transport), equation of state, etc., are implemented in
identical ways for both codes.  The two codes have already been
described in detail (Herant et al. 1994, hereafter HBHFC).  Thus, we
will limit ourselves to an overview except to describe modifications
introduced specifically for these simulations.

\subsection{One-dimensional Lagrangian Code}

The one-dimensional simulations were performed with an explicit,
grid-based, Lagrangian code (Benz 1991) using a second order
Runge-Kutta integrator. This code does not include any form of
convection modeling (mixing length or other), and as a result,
its usefulness is limited to non-convective regimes, and to the
diagnosis of the onset of convection.  The neutron star surface
is modeled by a reflective inner boundary, exterior to which lies the
atmosphere.  For most of the one-dimensional simulations we used a
total of 140 cells modeling the pressure equilibrium atmospheres out
to a radius $\sim$2500 km and the infall atmospheres out to
$\sim$25,000 km.  Since the mass of the atmosphere strongly depends
on entropy for the equilibrium models [see eq. (\ref{eq:matm})]
and on the infall rate for the infall models [see eq.
(\ref{eq:rff})], the mass resolution for the models varies according
to the computed model. It was however chosen in such a way as
to maximize resolution near the surface of the neutron star.

Both Newtonian and general relativistic (in the style of
Van Riper 1979) formalisms have been implemented in the code. We
have found that the general relativistic implementation leads
to an increase in the Brunt-V\"{a}is\"{a}la frequency of up to
$40 \%$ over the Newtonian case in the intermediate entropy
models (see \S \ref{sec:cei}), and in addition, a factor
of $\sim 3$ increase in accretion rates.  These effects are
comparable to those calculated by HC. However, since angular
momentum and accretion concerns limit us to mostly qualitative
estimates anyway, in most of the simulations presented here, general
relativistic effects have been ignored.

\subsection{Two-dimensional SPH Code}

The basic structure of the two-dimensional cylindrical geometry
smooth particle hydrodynamics (SPH) code used for the simulations
of this paper has been presented in Herant \& Benz (1992).
The code was further developed in Herant, Benz, \& Colgate
(1992) and HBHFC to incorporate neutrino processes
and an equation of state of extended range.

As in HBHFC we run our calculations in a wedge centered on the
equatorial plane with periodic boundary conditions to avoid the
complications associated with the $z$-axis which corresponds to
a singularity of the cylindrical coordinate system. The opening
angle is usually $90^\circ$ corresponding to $\sqrt2/2$ of the
total volume. We have varied the opening angle (up to $160^\circ$)
and the number of particles [factors of 4 (4,000 - 16,000 particles)]
in the simulations without noticing appreciable changes in the
results.  The gravitational force is calculated in the Newtonian
limit by evaluating the mass interior to each particle, thereby
assuming spherical symmetry, which is reasonable considering
that the central neutron star provides the dominant contribution.
The kernel used is the same as in HBHFC.

The structure of the initial atmospheres (see \S \ref{sec:ini})
is mapped to an SPH representation. Particles are placed on
concentric circles around the origin. There are of order 50
particles per circle, which in a $90^\circ$ wedge translates
to an angular resolution of a few degrees.  As in the
one-dimensional case, the mass of the particles is dependent
upon the specific characteristics of the atmosphere being
studied with the maximal mass resolution near the surface of
the neutron star.  Also similar to the one-dimensional code, a fixed,
hard boundary represents the neutron star surface.  Its
implementation is as described in HBHFC. Once particles reach
a critical density ($\rho > 10^{13}$ g cm$^{-3}$) and electron
fraction ($Y_e < 0.1$), they are accreted onto the neutron star
surface.  In addition, an outer boundary was introduced to
allow us to control the pressure of the outer atmosphere.

\section{INITIAL CONDITIONS AND PHYSICAL PROCESSES}

\subsection{Initial Conditions}\label{sec:ini}

In most of our simulations, the neutron star has a 1.4 $M_{\odot}$
gravitational mass and a radius of $10$ km.  Atmospheres with
initially uniform entropy and chemical composition are constructed
above the neutron star.  Their initial density structure is
determined by pressure equilibrium in the equilibrium atmosphere
models, or by assigning a mass infall rate and assuming free-fall
initial conditions for the infall simulations.  We considered
initial compositions of pure iron, pure oxygen, and a primordial
mix of hydrogen and helium. Because the temperature at the base
of the atmosphere is usually high enough to lead to nuclear
statistical equilibrium (NSE), the initial composition is
important to determine the energy release from nuclear burning.

For the equilibrium atmosphere calculations, pressure equilibrium
was verified in our code by allowing the atmosphere to evolve
hydrodynamically without neutrino physics. For the two-dimensional
simulations, inaccuracies in the mapping scheme from the
one-dimensional structure lead to small initial transients which
have to be damped, preserving constancy of entropy. In general
however, the atmospheres were found to remain in equilibrium for
times much longer than our simulation times.

\subsection{Equation of State and Neutrinos Processes}

The equation of state and neutrino processes are discussed in
HBHFC to which the reader is referred for further details.  The
equation of state includes perfect gas, photon, and electron
contributions to any degree of degeneracy and relativism (Nadezhin
1974; Blinnikov, Dunina-Barkovskaya, \& Nadezhin 1995), and a
nuclear equation of state (Lattimer \& Swesty 1991).
Additionally, when the temperature rises above $5\times10^9$ K,
NSE (see Hix et al., 1995) is enforced to approximate the effect
of nuclear burning, and to compute the free nucleon fraction which is
important for neutrino emission and absorption. Although matter does
burn at lower temperatures without going immediately into NSE,
burning is a slow process compared to the hydrodynamics.

The neutrino emission processes accounted for include electron
and positron capture by free protons and neutrons, and pair and
plasma neutrino--antineutrino creation. Neutrino absorption
processes include electron neutrino capture by free neutrons,
electron antineutrino by free protons, and neutrino antineutrino
annihilation. Neutrino scattering includes electron and positron
scattering of neutrinos, and neutral current opacities by nuclei.
Three species of neutrinos are tracked separately by the transport
algorithm: electron neutrino, electron antineutrino, and a generic
``$\tau$'' neutrino bundling together $\mu$ and $\tau$ neutrinos
and antineutrinos which have very similar characteristics in the
regimes under consideration.  The neutrino transport consists of
two schemes:  flux-limited diffusion for the optically thick regions
and a light-bulb approximation for the optically thin regions.  The
light-bulb approximation was introduced for computational speed and
is only valid if the absorbed neutrino luminosity ($L_{\nu_{\rm abs}}$)
in this regime is much less than the total neutrino luminosity
($L_{\nu_{\rm tot}}$).  In most of our calculations, we limit the
$L_{\nu_{\rm abs}}/L_{\nu_{\rm tot}}$ to $10 \%$.  We have run test
calculations limiting $L_{\nu_{\rm abs}}/L_{\nu_{\rm tot}}$ to $3 \%$
without appreciable changes in the results.

\subsection{Infall rates and Assumptions}

The objective of this paper is to determine the effects of
neutrino processes on accretion for a range of initial conditions.
We then apply our results to specific circumstances which lead to
rapid infall on neutron stars.  In this section, we discuss the
probable infall rates for various scenarios and the suitability
of our assumptions, which include ignoring the effects of angular
momentum, magnetic fields, and photon diffusion.

\subsubsection{Infall}

The infall of matter onto a neutron star plowing through a medium
is characterized for many, if not all, of the formation scenarios
by Bondi-Hoyle ``infall'' and has been further studied numerically
in a series of papers by Ruffert (1994a, b, 1995) and Ruffert \&
Arnett (1994).  They compare numerical results to the
canonical Bondi-Hoyle infall in a homogeneous medium, and
introduce a new set of equations to estimate the infall rate.
However, except at Mach numbers ($M$) close to 1, the
infall rate is within 20 $\%$ of the canonical rate (it can
increase by factors of 3 for $M=1$ infall).  Since we are only
interested in rough estimates of the infall, we will use the
simpler equations of Bondi-Hoyle-Lyttleton (Bondi 1952):
\begin{equation}
\dot{M}_{\rm B} \approx 4 \pi r_{B}^{2} \rho (v^{2} + c_{s}^{2})^{1/2}
\end{equation}
where the Bondi infall radius is
\begin{equation}
r_{B} = \frac{G M_{\rm NS}}{v^{2}+c_{s}^{2}},
\end{equation}
and $G$ is the gravitational constant, $\rho$ and $c_{s}$ are
the density and sound speed of the ambient medium.  Table 1
lists the infall radii and infall rates for a neutron
star plowing through stars (modeled with a stellar structure
code developed by D. Arnett) of different masses and at
different distances from the center for a range of impact
velocities spanning possible high infall scenarios.
Note that proper motion measurements of neutron stars imply
an average spatial velocity of neutron stars on the order of
$450$ km s$^{-1}$ and, in some cases, as high as $1000$ km s$^{-1}$
(\cite{Lyn94,Fra94}).  However, common envelope systems will
involve much lower velocities ($\sim v_{\rm orbital}$).

\subsubsection{Angular Momentum}\label{sec:ang}

Estimating the effects of inhomogeneous media on Bondi-Hoyle
accretion has been fraught with difficulties both from an
analytic and numeric standpoint.  Analytical approaches tend
to depend heavily upon unphysical assumptions (sometimes assuming
that the fluid flow system is identical to the standard
Bondi-Hoyle structure for the homogeneous case) and in general,
results have not been corroborated by simulations.  Moreover,
the lack of agreement among numerical endeavors has made their
estimates equally inconclusive.  A partial list of numerical
studies representing the large variation in results includes:
Davies \& Pringle (1980); Fryxell \& Taam (1988); Taam \&
Fryxell (1989); Sawada et al. (1989); Theuns \& Jorissen (1993);
and Ruffert \& Anzer (1994). Most of these discrepancies are
probably due to differences in resolution, boundary effects, or a
difference in accretion between two- and three-dimensional
models.  However, the numerical simulations virtually all
agree on two points:  the average infall rate is within a
factor of two of the Bondi-Hoyle infall (even though in
many simulations the infall rate is seen to vary with time,
e.g. the ``flip-flop'' instability seen by Fryxell and Taam), and
the angular momentum accreted within the accretor radius (typically
$0.1 r_{B}$) is generally non-zero, but less than that
predicted by a majority of the analytical estimates.

Despite these difficulties, we would like to try to evaluate,
using currently favored numerical models, the effects of
angular momentum on accretion.  A typical estimation for angular
momentum accretion is (see, for example, Ruffert \& Anzer 1994):
\begin{equation}
j_z \equiv \dot{J_z}/\dot{M} = \frac{1}{4} (6 \epsilon_{v} -
\epsilon_{\rho}) V r_{B} \label{eq:ang}
\end{equation}
where $\dot{J_{z}}$ is the angular momentum accretion rate,
$\dot{M}$ is the mass accretion rate and $V$ is the velocity of
the accretor, and  $\epsilon_{v,\rho}$ are the inhomogeneity
parameters defined as (Taam \& Fryxell 1989):
\begin{equation}
\epsilon_{\rho,v} = r_{B}/H_{\rho,v}
\end{equation}
where $H_{\rho,v}$ is the scale height of the density ($\rho$) or
the velocity (including the sound speed, i.e. $(c_s^2+v^2)^{1/2}$)
profile perpendicular to the direction of motion of the accretor.
Ruffert and Anzer consider a $V = 3 c_s$ accretor in a
$\epsilon_{v} = 0.3$ medium with an accretion radius one tenth
the size of the Bondi radius.  Their results give a $40 \%$
decrease in the angular momentum accreted in comparison to
equation (\ref{eq:ang}).  Assuming no angular momentum is lost
once the material passes within the accretion radius and using
equation (\ref{eq:ang}), one can determine an upper limit for the
radius at which rotational support stops the infall ($r_{\rm ang}$)
of material (also shown in Table 1). This is most likely an
over-estimate since, until an axisymmetric regime is attained,
angular momentum can be effectively transported by pressure
waves and shocks.

{}From $r_{\rm ang}$ inward, a thick accretion disk forms.
For a thick $\alpha$ disk, the timescale for the outward
transport of angular momentum is estimated to be (Chevalier 1993):
\begin{equation}
t_{\rm in} \sim \frac{r_{\rm ang}^2}{\alpha c_s H} \sim
\frac{r_{\rm ang}^{3/2}} {\alpha \sqrt{G M}} \sim r_8^{1.5}\quad {\rm s}
\end{equation}
where the sound speed ($c_s$) is approximated as the orbital
velocity, H is the disk scale height and is estimated to be
$\sim r$, $\alpha$ is a measure of the viscous stress typically
found to be $\sim .01-.1$ (we use .07), and $r_8$ is $r_{\rm ang}$
in units of $10^8$ cm. Once the material loses its angular
momentum, it will accrete onto the neutron star.  Thus $t_{\rm in}$
is essentially the time delay for the accretion of material.
Since $t_{\rm in}$ is not much larger than the free-fall
timescale, this delay should have a relatively small impact
on the accretion rate.

Despite this discussion, we are aware that the question
of the effect of angular momentum accretion is unsettled, and
it will remain so until more comprehensive results from
numerical and analytical calculations become available.
However, we believe that in the case of buried neutron stars,
the effect of angular momentum should be a perturbation
on the classical Bondi regime. One reason is that the dynamics
of self-gravitating, thick accretion disk is ripe with
hydrodynamical instabilities (unlike cold, thin disks), which
should allow rapid transport of angular momentum. Another
reason for sustained accretion is that just as there exists a
plane in which the angular momentum accreted is maximal, there
is also a plane in which it is nil, and that is where the
accreted matter may come from primarily (i.e. the polar
direction in the case of an inspiral).

\subsubsection{Photon Diffusion}\label{sec:pho}

Our current set of simulations is limited by the assumption
that the photons are ``trapped'' within the material of the
atmospheres.  By trapped, we mean that the photons are
carried inward with the accretion flow significantly faster
than they can diffuse outward.  As a result, we only
consider cases for which the accretion rate is sufficiently
high (infall models), or for which the entropy is not too large
(equilibrium atmosphere models). We derive below the conditions
for which the assumption of photon trapping is valid.

For equilibrium atmospheres, the timescale for neutrino cooling
($\tau_{\nu}$) sets the accretion rate and therefore the dynamical
timescale. We can therefore compare the photon diffusion timescale
($\tau_{\gamma}$, calculated near the surface of the neutron star
since this is where the interesting dynamics occur) with $\tau_{\nu}$
to determine the range of constant entropy, equilibrium atmospheres
for which we can assume neutrinos are trapped.  Using the
approximations for the structure of the equilibrium atmospheres
derived in \S \ref{sec:eqatm}, we obtain:

\begin{equation}
\tau_{\nu} = E_{\rm reg}/L_{\rm reg} =
3.2 \times 10^{-12} S_{\rm rad}^6 (1 + S_{\rm rad}/55)
\; {\normalsize s}
\end{equation}
and
\begin{equation}\label{eq:phd}
\tau_{\gamma} \approx \frac{r_{\rm scale}^2}{\lambda_{\rm mfp}^2}
\frac{\lambda_{\rm mfp}}{c}
\end{equation}
where $E_{\rm reg}$ and $L_{\rm reg}$ are the energy and
neutrino luminosity near the surface of the neutron star over
the scale height $r_{\rm scale}$ [see equation (\ref{eq:acc})],
$S_{\rm rad}$ is the radiation entropy of the atmosphere,
$\lambda_{\rm mfp}$ is the mean free path of the photons, and
$c$ is the speed of light.  For entropies in which
$\tau_{\gamma} > \tau_{\nu}$, ($S < 600$ - see Fig. 1)
the photons will be carried in with the accreting material and
we can reliably assume that the photons are trapped.  However,
this argument implicitly relies upon a spherically symmetric
inflow.  When convection becomes important, we must once again
examine the effects of this convection-enhanced photon diffusion.
We will consider these specific cases as they appear in our
equilibrium simulations.

For the infall atmospheres, we can use the infall rate to
determine the inward motion of the material and compare it
to photon diffusion.  The trapping radius is then the
radius where these speeds are equal and is commonly
denoted (e.g. see Chevalier 1989):
\begin{equation} \label{eq:pht}
r_{\rm tr} = \min \left( \frac{\dot{M} \kappa}{4 \pi c},r_{B} \right)
\end {equation}
where $\dot{M}$ is the mass infall rate, $\kappa$ is the opacity
of the infalling material (we assumed $\kappa = 0.2$ cm$^2$ g$^{-1}$),
$c$ is the speed of light and $r_{B}$ is the Bondi accretion radius.
{}From Table 1, we see that the trapping radius for most neutron star
encounters with stellar objects is close or equal to the Bondi
radius and, as a result, the diffusion of photons is unimportant.

\subsubsection{Magnetic Fields}\label{sec:mag}

We have assumed in our simulations that the magnetic field of
the neutron star has no effect upon the hydrodynamics.  This
assumption is valid for two regimes; for neutron stars with low
magnetic fields and for high infall rates which smother the
magnetic field of the neutron star.  The importance of magnetic
fields can be estimated by
finding the radius at which the magnetic energy density equals
the kinetic energy density of the infalling matter (Shapiro \&
Teukolsky 1983):

\begin{equation}
\frac{B^{2}}{8\pi} = \frac{1}{2} \rho v_{\rm ff}^{2}
\end{equation}
\begin{equation}
\Longrightarrow \frac{\mu^{2}}{8 \pi r_{A}^{6}} = \frac{1}{2} \frac{\dot{M}}
{4 \pi r_{A}^{2} v_{\rm ff}} v_{\rm ff}^{2}
\end{equation}
Where $B$ is the magnetic field, $\mu$ is the magnetic dipole moment,
$\rho$ is the density of the matter, $v_{\rm ff}$ is the free-fall
velocity, $r_{A}$ is the Alfven radius and $\dot{M}$ is the mass
infall rate.  To insure that magnetic fields are unimportant, the
Alfven radius should lie within the neutron star. For a $10^{12}$
Gauss magnetic field, the accretion rate has to be:

\begin{equation}
\dot{M} \gtrsim
0.8 \; M_{\odot} \; {\rm y}^{-1}.
\end{equation}
Thus for high infall rates the infalling material effectively
smothers even high magnetic fields and its trajectory is unaffected by
them.

However, the above calculation may be overly conservative because
it ignores the pile-up of material which occurs at lower
infall rates. As the infalling material builds up around the
neutron star, the pressure at the base of the atmosphere increases.
In a more detailed analysis, Chevalier (1989) calculated the
relative importance of radiation and magnetic pressure for this
built up material around a neutron star.  He found that for
the radiation pressure to exceed the magnetic pressure at
the surface of a $10^{13}$ Gauss neutron star, the infall rate
need only exceed $5 \times 10^{-6} M_{\odot}$ y$^{-1}$, in
support of our assumption that magnetic fields are unimportant
at high accretion rates.

\section{RESULTS}

Following the strategy that we mapped out in the introduction,
we present two sets of simulations for which the initial atmosphere
structures were defined by either free-fall or pressure equilibrium
conditions.  Recall that the infall atmospheres model the likely
initial conditions encountered by a neutron star entering a medium.
The infall simulations provide clues to the structure of the
developing atmosphere which, in most cases, will become similar to
our second set of models, the equilibrium atmospheres.  From the
equilibrium atmosphere simulations, we can then determine the ultimate
fate of these systems.  In both cases, we found that the outcome is
primarily dependent upon one parameter:  the infall rate for the infall
atmospheres and the entropy for equilibrium atmospheres.
In this section, we also compare our results to prior analytical
derivations obtained by C89 and HC for infall atmospheres and CHB
for equilibrium atmospheres.

\subsection{Infall Atmospheres} \label{sec:inf}

The structure of infall atmospheres has been discussed analytically
by C89 and in more detail by HC and Brown (1995).  The qualitative
structure, and indeed much of the quantitative results, predicted by
C89 is in good agreement with our numerical simulations.  The density
and pressure profiles derived by C89, with only a slight modification
of the adiabatic index, match our numerical simulations closely.
Still, as we shall see further, the seemingly innocuous deviation of
the adiabatic index for very high infall rates has non-negligible
consequences on the overall hydrodynamical evolution.  Yet of greater
importance, convection adds a new dimension to the problem as
high-entropy bubbles drive the accretion shock outward.  We begin
this section with a summary of conditions prevailing in steady state
infall atmospheres, much along the ideas set forth by C89.  We then
go on to the analysis of the initial transient which occurs during
the onset of accretion and has important consequences for the
subsequent evolution.  We then discuss our one-dimensional simulations
and compare them to the work of C89.  We end this section discussing
the effects of convection on infall atmospheres.

\subsubsection{Steady State Infall Structure}

As a neutron star plows through an ambient medium, material within
the Bondi infall radius is captured and falls inward.  Initially,
pressure forces are insignificant so that the infall structure is
similar to that predicted by the free-fall solution.  Rapidly
however, ``the sink backs up'' as the Bondi infall  rate
[$\dot M_B = 4 \pi r_B^2 \rho_{\rm ext} (c_{\rm ext}^2+v_{\rm NS}^2)^{1/2}$
where the ``ext'' subscript indicates the external density and sound speed]
exceeds the acceptance rate determined by the cooling rates.
As matter continues to crash down, it heats up and pressure increases
until it is sufficient to slow down the infall by a shock.
The accretion shock moves outward from the neutron star, until
the flow structure (especially conditions in the neighborhood
of the neutron star) allow neutrino losses to match the
energy output due to accretion, or in the case of low infall
rates, until the shock emerges from the optically thick domain.
In the latter case, which we will not consider further, the
accretion is  Eddington-limited, while in the former case,
neutrino cooling allows an essentially unlimited accretion rate.

The accretion shock separates two distinct structural regions in
the infall (see Fig. 2).  Outside the shock, the infall can
be characterized by the free-fall solution (see, for example C89):
\begin{equation}
v_{\rm ff} = \sqrt{\frac{2GM}{r}} \label{eq:vff}
\end{equation}
and
\begin{equation}
\rho_{\rm ff} = \frac{\dot{M_B}}{4 \pi r^2 v_{\rm ff}}, \label{eq:rff}
\end{equation}
where $G$ is the gravitational constant, $M$ is the neutron star
mass and $r$ is the distance from the center of the neutron star.

At the shock, using mass, momentum and energy conservation in
addition to a perfect gas equation of state and assuming a strong
shock, the shock jump conditions are (C89):
\begin{equation}
P_{\rm sh} = \frac{\gamma + 1}{2} \rho_{\rm ff} v_{\rm ff}^2 \label{eq:psh}
\end{equation}
and
\begin{equation}
\rho_{\rm sh} = \frac{\gamma + 1}{\gamma - 1} \rho_{\rm ff} \label{eq:rsh}
\end{equation}
where the pressure outside the shock is considered to be negligible
(strong shock assumption) and $\gamma$ is the adiabatic index.

Inside the shock front, in the settling region, we again turn
to the conservation equations assuming spherical symmetry and
that all initial transients set by the outmoving shock wave
are quickly ironed out by convection (C89):
\begin{equation}
\partial \rho/\partial t + \nabla(\rho v) = 0 \qquad
{\rm \normalsize mass \; conservation},
\end{equation}
\begin{equation}
\partial v/\partial t + \nabla (v^2/2 + P/\rho) = F_T
\qquad {\rm \normalsize momentum \; conservation}, \label{eq:momcon}
\end{equation}
and
\begin{equation}
P \propto \rho^{\gamma}
\qquad {\rm \normalsize adiabatic\; perfect \; gas},
\end{equation}
where $F_T$ is the net external force (in our case, gravity).
Steady state solutions demand that
$\partial v/\partial t=\partial \rho/\partial t=0$,
so that the mass conservation equation becomes
$4\pi r^2 \rho v = {\rm const} =\dot M_B$. Moreover,
right behind the shock radius, we have that
$v^2/2 \ll P/\rho$ since the flow becomes subsonic
($v^2 < c_s^2=\gamma P/\rho\propto \rho^{\gamma-1}$). From
the mass conservation equation, $v^2\propto \rho^{-2} r^{-4}$
so that the ratio $(v^2/2)/(P/\rho)\propto r^{-4}\rho^{-1-\gamma}$.
Since $\rho$ is obviously a decreasing function of radius,
we have that $v^2/2 \ll P/\rho$ everywhere behind the shock.
One can thus neglect the $v^2$ term in the momentum conservation
equation (\ref{eq:momcon}) throughout the region behind the
shock, so that the determination of the density and pressure
becomes straightforward (C89).
\begin{equation}
\rho = \rho_{\rm sh} \left( \frac{r}{r_{\rm sh}} \right) ^{-1/(\gamma-1)}
\end{equation}
and
\begin{equation}
P = P_{\rm sh} \left( \frac{r}{r_{\rm sh}}\right)^{-\gamma/(\gamma-1)}.
\label{eq:pgam}
\end{equation}
As the pressure, and hence temperature, at the base of the
atmosphere increases, the neutrino emission increases.
Because of the low cross-section of interaction with matter
($\sigma_\nu=10^{-43}$ cm$^2$ vs.
$\sigma_\gamma=\sigma_T=6\times10^{-25}$ cm$^{2}$),
the neutrinos are not trapped
like the photons and commence cooling the base of the atmosphere.
An important cooling mechanism is the capture of electrons and
positrons by free protons and neutrons with emission of electron neutrinos
and anti-electron neutrinos respectively. In the regime where
pairs dominate and the matter is completely dissociated into
free nucleons (corresponding to a high temperature and entropy),
the emission rate can be approximated by (Herant et al. 1992):
\begin{equation}
\frac{d\epsilon_{\rm nuc}}{dt} = 2\times10^{18} T^6_{\rm MeV}
\quad {\rm \normalsize erg \; g^{-1} \; s^{-1}}. \label{eq:nuc}
\end{equation}
Also important is the annihilation of electrons and positrons
into neutrino antineutrino pairs of all flavors which, if
pairs dominate, can be written as (Herant et al. 1992):
\begin{equation}
\frac{d\epsilon_{\rm pp}}{dt} = 1.9\times10^{25}
\frac{T^9_{\rm MeV}}{\rho}
\quad {\rm \normalsize erg \; g^{-1} \; s^{-1}}. \label{eq:pp}
\end{equation}

In steady state, the neutrino losses balance the
gain in potential energy due to accretion;
\begin{equation}
4 \pi r_{\rm NS}^2 (\Delta r_{\rm ER}) \rho \frac{d\epsilon_{\rm tot}}
{dt} = \frac{G M_{\rm NS} \dot{M_B}}{r_{\rm NS}} \label{eq:insh2}
\end{equation}
where $\Delta r_{\rm ER}$ is the thickness of the emission region
at the base of the atmosphere where most of the cooling takes
place (one temperature scale height, $\sim r_{\rm NS}/8$),
$d\epsilon_{\rm tot}/dt$ is the specific neutrino cooling rate
from the neutrino emission processes[eqs. (\ref{eq:nuc}),
(\ref{eq:pp})].  This last expression closes
the set of equations that determines
the steady state of the system and allows one to determine
$r_{\rm sh}$. For instance, we can use the approximation
that the pressure is dominated by radiation at the base
of the atmosphere (i.e. $P_{\rm atm}=a T^4_{\rm atm}$, $\gamma=4/3$)
to determine the temperature at the base of the atmosphere from
equations (\ref{eq:vff})-(\ref{eq:pgam}), and
thus the cooling rate for free nucleon emission:
\begin{equation}
\frac{d\epsilon_{\rm nuc}}{dt} = 8.5 \times 10^{15}
\left( \frac{M}{1.4\, M_\odot} \right)^{3/4}
\left( \frac {\dot{M_B}}{M_\odot {\rm y}^{-1}} \right)^{3/2}
\left( \frac{r_{\rm NS}} {10\,{\rm km}} \right)^{-6}
\left( \frac{r_{\rm sh}}{100\,{\rm km}} \right)^{9/4}
\quad {\rm \normalsize erg \; g^{-1} \; s^{-1}},
\end{equation}
and pair annihilation:
\begin{equation}
\frac{d\epsilon_{\rm pp}}{dt} = 9.2\times10^{16}
\left( \frac{M}{1.4\, M_\odot} \right)^{13/8}
\left( \frac{\dot{M_B}} {M_\odot {\rm y}^{-1}} \right)^{5/4}
\left( \frac{r_{\rm NS}}{10\,{\rm km}} \right)^{-6}
\left( \frac{r_{\rm sh}}{100\,{\rm km}} \right)^{15/8}
\quad {\rm \normalsize erg \; g^{-1} \; s^{-1}}.
\end{equation}
For most steady-state accretion scenarios, the pair annihilation
emission process dominates the cooling, so we will ignore nucleon
emission in determining $r_{\rm sh}$.  Assuming a $10$ km,
$1.4\, M_\odot$ neutron star, we obtain a result similar to C89:
\begin{equation}
r_{\rm sh}^{\gamma = 4/3} = 6.7\times10^8 {\dot{M_B}}^{-10/27} \, {\rm cm},
\label{eq:rsh1}
\end{equation}
where $\dot{M_B}$ is in $M_\odot$ y$^{-1}$.
Our neutrino emission processes are slightly different from
those in C89 which, combined with a different value for the
emission region (we used $r_{\rm NS}$, accounts for the slight
differences between the two solutions.  We can repeat this
derivation for various values of gamma to obtain:
\begin{equation}
r_{\rm sh}^{\gamma = 1.37} = 3.3\times10^9 {\dot{M_B}}^{-0.46} \, {\rm cm}
\end{equation}
and
\begin{equation}
r_{\rm sh}^{\gamma = 1.4} = 1.7\times10^{10} {\dot{M_B}}^{-0.56} \, {\rm cm}.
\label{eq:rsh3}
\end{equation}

\subsubsection{The Onset of Accretion and the Initial Transient}

We now leave the steady state solution and return to the
analysis of the initial transient which takes place
at the beginning of the evolution. During the initial
phase of the accretion, the shock moves out from the
surface of the neutron star towards its steady-state
position $r_{\rm sh}$.  As a result, the post-shock entropy
decreases as the accretion shock progresses outwards
and weakens.  Assuming that radiation dominates, but that
electrons are non-relativistic (which is reasonable
away from direct proximity of the neutron star), the
post-shock entropy can be written
$S_{\rm sh}=1.1 \times 10^{-11} P_{\rm sh} ^{3/4} / \rho_{\rm sh}$.
We can then apply the shock equations
(\ref{eq:psh}) and (\ref{eq:rsh}) and the free-fall equations
(\ref{eq:vff}) and (\ref{eq:rff}) to determine the entropy
for a given shock radius and infall rate:
\begin{equation}
S=5.5 \times 10^{3} \frac{((\gamma+1)/2)^{3/4}}
{(\gamma+1)/(\gamma-1)} M_{\rm NS}^{7/8} \dot{M_B}^{-1/4} r_{6}^{-3/8}
\label{eq:sin}
\end{equation}
where $S$ is in $k_B$ per nucleon, $M_{\rm NS}$ is the mass of the
neutron star in solar masses, $\dot{M_B}$ is the infall rate
in $M_{\odot}$ y$^{-1}$, and $r_6$ is the radius of the shock in
units of $10^6$ cm.  Note that in the strong shock regime (which
applies if the shock radius remains much smaller than the
Bondi radius), the entropy below the shock is independent of
the entropy above the shock.  Consequently, in most of our
simulations, the initial entropy of the infalling material
has little effect on the resulting structure.

Because of the dependence on radius (exponent $-3/8$)
of the postshock entropy, the outward motion of the shock
imprints a negative entropy gradient in the inner region
during the initial transient.  This entropy profile is
evidently unstable and leads to a break in the spherical symmetry
which is not included in the picture developed in C89.
The timescale for convection can be approximated using the
Brunt-V\"{a}is\"{a}la frequency $N$ (see Cox, Vauclair,
\& Zahn 1983):
\begin{equation}\label{eq:brv}
N^2 = \frac{g}{\rho} \Bigl(\frac{\partial \rho}{\partial S}\Bigr)_P
\frac{\partial S}{\partial r}.
\end{equation}
For $\partial S/\partial z < 0$, $N^2 < 0$ and the atmosphere is
unstable.  The timescale for this convection ($\tau_{\rm conv}$) is
$\sqrt{\left| 1/N^2\right|}$.  Of course, convective instability
can also depend on the chemical composition as an atmosphere may be
stabilized by a negative gradient of the molecular weight. However,
the entropy gradients are sufficiently high in our simulations that
this is not a concern.  The region
within the accretion shock convects until stability is
achieved and ultimately develops into an equilibrium
atmosphere.

An additional effect of this initial convection is to drive the
shock beyond its steady-state value, lowering the entropy at the
shock [eq. (\ref{eq:sin})] with respect to the steady-state
prediction.  This material ultimately makes its way down to the base of
the atmosphere, defining the entropy of the equilibrium atmosphere.
Since the base of the atmosphere depends on its entropy, this convection-
driven overshoot can drastically alter the evolution of the system.
Equilibrium atmospheres and the effect of the initial transient on
their entropy will be addressed in \S \ref{sec:eqatm}.

However, this description does not apply to extremely
high infall rates. Beginning from the free-fall solution
of matter crashing down on the neutron star surface we have:
\begin{equation}
\frac{{GM_{\rm NS}}{r_{\rm NS}}}{\rho_{\rm atm}}=aT_{\rm atm}^4
\end{equation}
so that the temperature at the bottom of the atmosphere is
proportional to the density at the bottom of the atmosphere
to the one-fourth power: $T_{\rm atm}\propto\rho_{\rm atm}^{1/4}$.
{}From mass conservation, we also have that
$\rho_{\rm atm}\propto\dot M$, and as a result
$T_{\rm atm}\propto \dot{M}^{1/4}$. Since neutrino cooling
has a high power dependence on temperature (see section
\ref{sec:eqatm}) there exists a critical mass infall
rate beyond which the accretion shock can hover right above
the neutron star and provide a sufficiently high temperature
for neutrino cooling to allow accretion to take place
immediately. In our simulations, this occurs at infall
rates upward of $10^5$ $M_\odot$ y $^{-1}$ (see Table 2).
The critical infall rates are slightly higher for the
higher initial entropy atmospheres due to their lower initial
neutrino luminosities.

Thus, rapid mass infall atmospheres lead to three distinct
regimes.  For a low rate of infall, material builds up around
the neutron star and sends an accretion shock outward from the
core. The unstable negative entropy gradient inprinted by
the motion of the shock leads to instabilities.  The
low-entropy matter at the shock is convected downward onto
the neutron star, and eventually an equilibrium atmosphere
forms, with the range of outcomes discussed in section
\ref{sec:eqatm}.   Intermediate rates of infall initially
lead to similar situations. However, the intense neutrino
emission leads to cooling times shorter than the convective
time, thus preventing the formation of an equilibrium
atmosphere.  For a high rate of infall, the material is shocked
at close proximity of the neutron star surface and cools
efficiently through neutrino emission without further ado.

\subsubsection{One-dimensional Simulations} \label{sec:in1d}

We have modeled a series of infalls for a range
of infall rates and initial entropies.  The results from the
one-dimensional simulations are shown in Table 2.  Note that
the initial entropy has little impact, except in mass accretion
rates near the transition between high and intermediate infall
regimes.

Although the structure of the infall atmosphere is reasonably
well described by the analytical derivation of C89, we found
that the effective adiabatic index $\gamma$ from our more detailed
equation of state is slightly higher than the radiation dominated
$4/3$ assumed by C89 (see Fig. 3).  For infall rates below
1 $M_{\odot}$ y$^{-1}$, this deviation is too small to have a
significant impact on the flow structure, but for higher infall
rates, it has a crucial influence on the steady state
position of the shock.  Figure 4 shows $r_{\rm sh}$ versus
accretion rates for a range adiabatic indices (1.33, 1.37, 1.40)
using equations (\ref{eq:rsh1}-\ref{eq:rsh3}).  Note
that by merely changing $\gamma$ from $4/3$ to $4.2/3$ for the
100 $M_{\odot}$y$^{-1}$ simulation changes $r_{\rm sh}$ by an
order of magnitude.  This is important because, as we have seen
in the previous section, the position where the shock stalls
determines the entropy of the atmosphere which develops in the
inner region in the vicinity of the neutron star. As we shall
discuss in section \ref{sec:tds}, this entropy determines the
ultimate fate of the atmosphere.

Figure 5 shows the entropy profile established by the transient
motion of the shock from the surface of the neutron star toward
its steady state radius. As predicted by equation (\ref{eq:sin}),
the entropy gradient is negative and will thus be subject to
convective instabilities. Table 2 lists Brunt-V\"{a}is\"{a}la
timescales along with neutrino luminosities, effective $\gamma$'s,
and central entropies for all the one-dimensional simulations.
It is clear, however, that multidimensional simulations are
needed to calculate the subsequent evolution.

\subsubsection{Two-dimensional Simulations}\label{sec:tds}

Since the high infall regime is not conducive to instabilities,
we have limited our two-dimensional simulations to intermediate
and low rates of infall.  These atmospheres become active quickly
(recall the short Brunt-V\"{a}is\"{a}la timescales) sending bubbles
outward through the inner region.  These bubbles
contribute to the outward push of the shock, while plumes of
low-entropy material stream down towards the neutron star
(see Fig. 6).  To better appreciate the properties of infalling
atmospheres, the distinction between intermediate and low infall
atmospheres needs to be elucidated.  In addition, for the low infall
regimes, we would like to determine the entropy of the resulting
equilibrium atmosphere.
As stated in the introduction, due to the relatively long convecting
timescales ($> 10^4$ s), it is impossible to simulate a complete
convective turn over as the Courant time step restriction near
the neutron star is of order of tens of $\mu$s. Nevertheless,
the simulations provide sufficient indications about the long
term behavior of the system to allow us to predict the ultimate
outcome of the evolution.

In our simulations, we have noticed that the typical velocity of an
infalling plume of material travelling between the shock and the
neutron star is $\sim 0.1$ the free-fall velocity.  Knowing this,
we can estimate the convective turnover timescale (see Table 3):
\begin{equation}
\tau_{\rm conv} = 10\,t_{\rm dyn} = 10\frac{\pi r}{c_s} =
10\frac{\pi r^{1.5}}{\sqrt{G M_{ns}}}\label{eq:tur}
\end{equation}
where $t_{\rm dyn}$ is the dynamical timescale, $r$ is the radius
of the material to be convected inward; $\tau_{\rm conv}$ is
the time to advect material from the shock to the neutron
start surface.  The time $\tau_{\rm conv}$ must be compared
with the time $\tau_{\rm BH}$ required for the neutron star to
accrete enough material to collapse into a black hole
(see Table 3):
\begin{equation}
\tau_{\rm BH} = \frac{1}{L_{\nu}} \frac{G M_{\rm NS} m}{r_{\rm NS}}
\end{equation}
where $L_{\nu}$ is the total neutrino emission per unit time,
$m$ is the additional mass required to induce collapse (we
use $0.2 \; M_{\odot}$), and $r_{\rm NS}$ is the neutron star radius.
When $\tau_{\rm conv} > \tau_{\rm BH}$ which corresponds to the
case of intermediate rate of infall, we expect the neutron star
to collapse into a black hole before an equilibrium atmosphere
can be formed.  However, when $\tau_{\rm conv} < \tau_{\rm BH}$,
which corresponds to low rates of infall, there is sufficient time
for convection to form an equilibrium atmosphere.  In these cases,
we can use equation (\ref{eq:sin}) to determine the resultant
entropy for the equilibrium atmosphere.

These results can be approximately summarized as follows. Very high
rates of infall ($\dot{M} \gtrsim 10^6\, M_{\odot} {\rm y^{-1}}$ --
Note that these numbers are estimates as these rates depend on additional
factors such as the Bondi radius and initial atmosphere entropy)
clamp the shock close to the neutron star and lead to rapid accretion
and black hole formation.  An intermediate rate of infall
($10^{3}\, M_{\odot} {\rm y}^{-1}$ $\lesssim \dot{M}
\lesssim 10^6\, M_{\odot}\, {\rm y}^{-1}$) does not have the time to
form a proper atmosphere as it rapidly leads to collapse into a black
hole.  Low rates of infall ($\dot{M} \lesssim 10^{3}\, M_{\odot}
{\rm y}^{-1}$) allow sufficient time for an equilibrium atmosphere
to develop. The structure and fate of these equilibrium atmosphere
are discussed in the next section.

\subsection{Equilibrium Atmospheres} \label{sec:eqatm}

In the previous section, we have studied the initial development of
an accretion structure around a  neutron star encountering an
external medium.  We now turn to the ultimate fate of these systems
after convective stability has been achieved.  Given the condition
of initial pressure equilibrium, the most massive atmosphere that
can form stably above a neutron star is isentropic, where the
entropy is determined by whichever density and temperature is
chosen at the surface of the neutron star. More massive atmospheres
can be constructed with a negative entropy gradient, but they are
unstable to convection.  Because even a small negative entropy
gradient rapidly drives convection [see eq. (\ref{eq:brv}) for
convective timescale], stellar models always adjust themselves to
constant entropy or positive entropy gradient structures.  We expect
the same situation for atmospheres around neutron stars.  Consequently,
we believe that the most simple and appropriate way to parametrize
the set of the possible atmospheres is to use isentropic initial
conditions.

The analytical work of CHB has examined the characteristics of
constant entropy equilibrium atmospheres with the additional
assumption that the pressure and internal energy are dominated
by radiation and electron pairs contributions (that is,
$P = 11/12 a T^4$).  This remains valid at the base
of the atmosphere for entropies less than 400 $k_B$/nucleon
[see eq. (\ref{eq:tatm})] and larger than 30 $k_B$/nucleon.
The radiation component of the entropy
(in units of Boltzman factor per nucleon) can then be expressed (CHB):
\begin{equation} \label{eq:srad}
S_{\rm rad}=\frac{4}{3} \times \frac{11}{4} a T^3/(\rho k_B N_A)
= 1.4 \times 10^{-11} P^{3/4}/\rho =
5.2 \times 10^8 T^3_{MeV}/\rho 	 \label{eq:rad}
\end{equation}
This expression is valid when the entropy and temperature are high.
For high entropies, $S_{\rm rad} \approx S_{\rm tot}$, so that the
analytical derivations can be compared directly to our models with
increasing accuracy the higher the entropy.

Assuming constant entropy and a radiation pressure dominated
system, one can use the hydrostatic equation of pressure equilibrium
to derive the structure of the atmosphere (CHB).

\begin{equation}\label{eq:srd}
P = [\frac{1}{4} M_{\rm NS} G (S_{\rm rad}/S_0)^{-1}(1/r-1/r_1)+P_1^{1/4}]^4
\quad {\rm \normalsize dyne \; cm^{-2}},
\end{equation}
where $S_0 = 1.4 \times 10^{-11}$ $k_B$/nucleon with
$r_1$ and $P_1$ referring to the radius and pressure at the outer
boundary.  Implicit to this derivation is that the mass of the
atmosphere is negligible when compared to the neutron star mass.
As we shall see, for high-entropy atmospheres, this is a good
assumption.

In most cases, the radius of the outer boundary is sufficiently
large that the terms involving $r_1$ and $P_1$ can be neglected.
In the equilibrium simulations, we have included varying degrees
of boundary pressure (motivated in part by the computed pressure
of the infalling material from our infall simulations - \S
\ref{sec:inf}).  Thus, we have incorporated the full equations
in our comparison with the analytic solutions.  Nonetheless,
the basic structure of the atmospheres changes very little (less
than a factor of 2 for even the most extreme external pressures)
by ignoring the boundary conditions, so we will present the
structure equations in their simple,
$P_1 = 0,\; r_{1} \gg r_{\rm NS}$ form (CHB):

\begin{equation}
P_{\rm atm} = 1.83 \times 10^{35}\, S_{\rm rad}^{-4}\, r_6^{-4} \quad
{\rm \normalsize dyne\;cm^{-2}},
\end{equation}
and using equation (\ref{eq:srad}), we find,
\begin{equation}\label{eq:rha}
\rho_{\rm atm} = 3.9 \times 10^{15}\, S_{\rm rad}^{-4}\, r_6^{-3} \quad
{\rm \normalsize g\; cm^{-3}},
\end{equation}
\begin{equation}
T_{\rm atm} = 195\, S_{\rm rad}^{-1}\, r_6^{-1} \quad {\rm \normalsize MeV},
\label{eq:tatm}
\end{equation}
\begin{equation} \label{eq:mat}
M_{\rm atm} = 24.5\, S_{\rm rad}^{-4}\, \ln(r_{\rm max}/r_{\rm NS}),
\label{eq:matm} \quad M_\odot
\end{equation}
where $r_6$ is the radius in units of $10^6$ cm, and $S_{\rm rad}$
is the radiation entropy in units of $k_B$/nucleon. These
expressions assume a 1.4 $M_{\odot}$ neutron star,
like in our simulations.  The most significant consequence from
setting $P_1 = 0$ and $r_1 \gg r_{\rm NS}$ is in the total
mass of the atmosphere, $M_{\rm atm}$. Figure 7 plots $M_{\rm atm}$
as a function of entropy for an outer radius of the atmosphere of
$10^9$ cm and $10^{13}$ cm using equation (\ref{eq:mat}) in the
case of no external pressure (dashed lines) and additional
external pressure (solid lines). When present, the external
pressure was determined using our infall models (see \S
\ref{sec:inf}) to find the maximum realistic infall pressure
($P_{\rm infall} = \frac{1}{2} \rho_{\rm ff} v^2_{\rm ff}$)
at a given radius and entropy.  These lines represent the maximum
achievable mass for a {\it stable} atmosphere of the prescribed
radius and entropy.  Note that despite the fact that constant
entropy structures are the most massive stable atmospheres,
their masses tend to be small.  More massive atmosphere
would require a negative entropy gradient which would then
be convectively unstable.

Using equations (\ref{eq:nuc}), (\ref{eq:pp}) combined with the
structure equations for temperature and density, it is
apparent that the neutrino energy emission per gram falls off
roughly as $r^{-6}$, implying that most of the neutrino emission (and
hence cooling of material) occurs close to the neutron star.
We estimated the cooling rate by assuming a nearly constant
neutrino cooling over a scale height of the emission region
(subscript ``ER'') $r_{\rm ER} = r_{\rm NS}/8$ above the
neutron star and calculating the neutrino luminosity per
gram in this region.  The mass accretion is then:
\begin{equation}
\dot{M} = M_{\rm ER} L_{\rm ER} / E_{\rm ER} \label{eq:acc}
\end{equation}
where $M_{\rm ER}$ is the mass within a scale height of the
neutron star, $L_{\rm ER}$ is the neutrino emission from that
region using equations (\ref{eq:nuc}) and (\ref{eq:pp}), and
$E_{\rm ER}$ was chosen (somewhat arbitrarily) to be half the
potential energy $G M_{\rm NS} M_{\rm ER}/(r_{\rm NS}+r_{\rm ER})$
gained by the material falling from infinity.  Combining
these equations and using the $P_1 = 0$ atmosphere structure,
we obtain:

\begin{equation}
\dot{M} = 9.0 \times 10^{11} S_{\rm rad}^{-10}\, (1+S_{\rm rad}/55) \quad
M_{\odot} \; {\rm \normalsize s^{-1}}
\end{equation}

Neutrinos emitted at the base of of the atmosphere can be recaptured
and heat matter higher up.  This is especially important
for low- and intermediate-entropy atmospheres ($S < 50$).
Neutrino absorption by free nucleons gives rise to the following
heating term (Herant et al. 1992):
\begin{equation}
\frac{d\epsilon}{dt} = 4.8 \times 10^{32} \frac{L_{\nu}}{4\pi r^2}
T^2_{\nu} \quad {\rm \normalsize erg \; g^{-1} \; s^{-1}}
\end{equation}
where $L_{\nu}$ is the electron neutrino luminosity which is
due to neutrino emission at the base of the atmosphere, and
$T_\nu$ is the neutrino temperature which tends to be similar to
the matter temperature near the surface of the neutron star.
We have seen earlier that cooling is proportional to $r^{-6}$, while
it appears that heating is proportional to $r^{-2}$.  One therefore
expects that there exists a radius separating an inner region where
cooling dominates from an outer region where there is a net gain in
energy from neutrino processes. This is known in supernova circles as
the gain radius.

The preceding equations give an analytical picture of the dominant
physical processes involved for atmospheres in which photons are trapped.
Note that nuclear burning effects were ignored. In our simulations,
we have noticed that the initial chemical composition of the
atmospheres has little effect upon the end result.  As was seen
analytically, we find that the primary parameter characterizing
the atmospheres is entropy.  For a low-entropy atmosphere, an
immediate explosion is generated by the intense emission of neutrinos
and the resulting energy deposition just beyond the gain radius.
Intermediate ranges for entropy still exhibit noticeable effects
from neutrino heating which induce convection.  For these entropies,
two-dimensional simulations of the atmospheres are required to fully
investigate the hydrodynamical evolution.  For high-entropy atmospheres,
neutrino heating turns out to be unimportant, and hence there is no
convection.  However, neutrino cooling continues to determine the
accretion rates up to extremely high entropies.  These results are
summarized in Table 4, and presented in more details in the following
sections.

\subsubsection{Low-entropy Atmospheres}\label{sec:iexp}

Figure 1 shows the sound travel time ($\tau_{s}$) through the
neutrino emission region together with neutrino cooling and
photon diffusion time scales. For $S_{\rm rad} \lesssim 14$
$k_B$/nucleon, which corresponds to $S_{\rm tot} \lesssim 22$
$k_B$/nucleon for our simulations, we find that the neutrino
cooling time is faster than the sound travel time.  Because
of this, it is physically impossible to form such an atmosphere
in hydrostatic equilibrium.  It is thus unlikely that
equilibrium atmospheres with low entropies can exist.  Just
to see what would happen, we have constructed such atmospheres
in pressure equilibrium artificially maintained by ignoring neutrino
effects. These atmospheres lead to neutrino-driven explosions
as soon as neutrino processes are turned on. Despite the
fact that they are unphysical, low-entropy atmospheres illustrate
the importance of neutrino energy deposition beyond the gain
radius. This process also plays a critical role in the more
physical scenarios of supernova explosion, or in intermediate
entropy atmospheres which are discussed below.

\subsubsection{Intermediate-entropy Atmospheres} \label{sec:cei}

For atmospheres within a range of intermediate entropies
($30 \lesssim S_{\rm tot} \lesssim 50$), the sound crossing time
is much less than the neutrino cooling time ($\tau_{s} \lesssim
0.01 \tau_{\nu}$), allowing the formation of atmospheres in
pressure equilibrium.  However, neutrino deposition is still
sufficiently strong to heat the atmosphere just beyond the
gain radius and thus raise the otherwise constant entropy of
this region above the value of the rest of the atmosphere.
The resulting negative entropy gradient is unstable and
convection takes place.  In supernova simulations (HBHFC), this
convection increases the efficiency of neutrino heating leading
eventually to an explosion.  In the context of our simulations,
in order for the convection to be important, it must be able to
overcome the general advection inward resulting from the sharp
decrease in pressure support as material near the neutron star
surface is quickly cooled by neutrino emission. Or in simpler
terms, the bubbles have to rise faster than they are dragged
inward by the general accretion flow. We can estimate the
relative importance of these effects in the atmosphere by
comparing the Brunt-V\"{a}is\"{a}la and infall time scales
(see Table 4).  Evidently, multidimensional simulations are
required to correctly model these phenomena.

Figure 8 shows the effects neutrino heating for an $S_{\rm tot}=50$
atmosphere on the entropy profile in a one-dimensional
computation. While neutrino cooling rapidly decreases the
entropy at the base of the atmosphere near the neutron
star, neutrino absorption further up leads to the formation
of an entropy peak and an associated negative entropy gradient
which will drive convective instabilities. When the same
atmosphere is simulated in two dimensions, large scale convection
arises from the neutrino induced negative entropy gradient
as can be seen in Figure 9.  Within this entropy range, the
two-dimensional calculations resulted in the expulsion of the
atmosphere in what might be considered a ``mini-supernova''
(see Fig. 10).  Table 4 gives explosion energies for atmospheres
of different entropies.  The energies are much lower
than supernova energies primarily due to the low mass of the
atmospheres ($M_{\rm atm}(1000\,{\rm km}) \sim 10^{-5} -
10^{-3} M_{\odot}$).

Rather than trap the released gravitational energy near the surface
of the neutron star to be ultimately emitted in neutrinos, convective
bubbles transport the energy up through the atmosphere as they rise.
Such extensive convection raises the question whether our one-dimensional
analysis which showed photon diffusion to be negligible is still
valid.  However, given the short timescales required
for the explosion to develop in our simulations, photon diffusion
remains unimportant.  For example, in the worst case scenario of
our $S_{tot} = 50$ atmosphere after $0.5$ s (see Fig. 11),
we find from equations (\ref{eq:rha}) and (\ref{eq:phd}) that
photons at $2000$ km diffused less than $1.5$ km during the
course of the simulation, a fraction of the particle size
at that radius.

\subsubsection{High-entropy Atmospheres}

In the case of high-entropy atmospheres ($S_{\rm tot} \gtrsim 60$),
neutrino deposition has little effect upon the infalling atmosphere.
Neutrino emission, however, remains an efficient source of cooling
for all our simulations which extend up to $S_{\rm tot} = 125$.
In fact, neutrino losses dominate photon losses up to $S_{\rm tot}=600$
(Fig. 1).

Figure 12 plots the entropy profile at discrete time intervals
(80 ms) for a typical one-dimensional run ($S_{\rm tot}=80$).
The cooled, low-entropy matter consists primarily of neutrons
($Y_e \sim 0.1$) and has essentially become part of the neutron
star.  Note that for these high-entropy atmospheres, no entropy
``bump'' develops through neutrino heating.  Figure 13
shows the neutrino luminosities and mean energies vs. time.
The neutrino luminosity increases initially and then stabilizes,
indicating a constant rate of accretion for the duration of the
simulation.  In essence, the system has reached a steady state
in which the neutrino emission exactly balances the compression
work done by gravity on the gas settling on the neutron star.
In these conditions, hypercritical accretion is maintained
until the neutron star collapses into a black hole.

For each simulation, the constancy of the accretion rate
was verified by fitting the accreted mass vs. time with
a straight line (see Fig. 14 for a typical fit), and the
r.m.s. deviation was estimated.  Figure 15 presents a
comparison between the accretion rate calculated analytically
and numerically as a function of atmosphere total entropy.
The open symbols were plotted using a straight insertion
of the total entropy while the filled symbols were plotted
by computing the radiation entropy from the numerical
simulations and using this value in equation (\ref{eq:acc}).
The remarkable agreement between analytical and numerical
calculations (when the radiation entropy is used) shows
that our models are self-consistent. In addition, we note
that as entropy increases, the difference between $S_{\rm tot}$
and $S_{\rm rad}$ becomes smaller and smaller.  Together with
the good agreement with the CHB model, this allows us to
extrapolate the behavior of our atmospheres beyond the range
of entropies that we have simulated.

Figure 16 show the analytical accretion rates for high-entropy
atmospheres.  The lower and upper curves represent the cases
with no external pressure, and with external pressure
at 1000 km of $0.25 \%$ the pressure at the surface of the
neutron star respectively. Of course, in reality the external
pressure on the atmosphere is determined by the formation
mechanism.  However, the limits that we have chosen bound the
results from all our infall models (see section \ref{sec:inf})
and it is unlikely that any of the formation mechanisms will
produce atmospheres with external pressures beyond these limits.
Note also that photons are still trapped out to 100 times the
neutron star radius at an entropy $\sim 600 \; k_B$/nucleon,
which is over a factor of ten times higher than typical stellar
entropy values.

\subsection{Summary of Results} \label{sec:summ}

We are now in position to tie together our studies of infall
models with the behavior observed in our equilibrium atmosphere
simulation, to create a complete picture of the evolution of
rapid accretion onto neutron stars (see table 5).  As we know from section
\ref{sec:eqatm}, if the entropy is greater than $\sim 600$, then
the atmosphere is be stable over long time scales.  For
$600 < S < 50$, the atmosphere accretes hypercritically.
But for $50 < S < 30$, neutrinos heat the base of the
atmosphere, ultimately leading to explosions.  The mass accreted
before the explosion is $M_{\rm exp} = \dot{M} \tau_{\rm exp}$,
where $\dot{M} = (G M_{\rm NS}/r_{\rm NS})/L_{\nu}$ and
$\tau_{\rm exp} \sim \tau_{\rm conv}$.  These values are shown
in Table 3.  The final results for the objects in Table 1 are
listed in the last column.  Immediate collapse designates
atmospheres in the high or intermediate regimes, $S > 600$
atmospheres are stable, $600 < S < 50$ atmospheres suffer
delayed collapse, and $50 < S < 30$ atmospheres result in
explosions.

We will now summarize these results in terms of accretion rates.  Very
high rates of infall ($\dot{M} \gtrsim 10^6\, M_{\odot} {\rm y}^{-1}$)
clamp the shock close to the neutron star and lead to rapid accretion
and black hole formation.  Very low rates of infall ($\dot{M} \lesssim
10^{-4}\, M_{\odot} {\rm y}^{-1}$) allow the entropy to rise
to about 600 $k_b$/nucleon and form a stable atmosphere lasting
many dynamical times.  Low rates ($10^{-4}\, M_{\odot}
{\rm y}^{-1}$ $\lesssim \dot{M} \lesssim 0.1\, M_{\odot} {\rm y}^{-1}$)
of infall form a stable atmosphere in near pressure equilibrium which
nonetheless accretes hypercritically and lead to the eventual formation
of a black hole. Medium-low ($0.1\, M_{\odot} {\rm y}^{-1}$ $\lesssim
\dot{M} \lesssim 10^3\, M_{\odot} {\rm y}^{-1}$) rates of infall also
form a stable atmosphere in near pressure equilibrium but neutrino
heating eventually leads to an explosion rather than black hole
formation. Finally, an intermediate rate of infall ($10^{3}\, M_{\odot}
{\rm y}^{-1}$ $\lesssim \dot{M} \lesssim 10^6\, M_{\odot} {\rm y}^{-1}$)
does not have the time to form a proper atmosphere as it rapidly leads
to collapse into a black hole.

\subsection{Initial Transient Revisited}

We have implicitly assumed in \S \ref{sec:summ} that the initial
transients define the fate of the system.  However, if a system
evolves into a steady state, this steady state can be maintained
as long as any variations in the infall rate are sufficiently
slow that the atmosphere can adapt to the changes before vigorous
convection develops.  The ultimate fate of the system
is identical under the steady-state solution and the transient
solution for all regimes except the medium-low regime which results
in explosions.  We will thus limit this discussion to the specific
cases where a steady-state system evolves and then the infall
rate is gradually changed to place it in the medium-low regime.
If the steady state is maintained, the
boundary between low and medium-low regimes will rise, limiting
the range of atmospheres that fall into the medium-low regime.
In this section, we will estimate how
slow the infall rate must change to maintain the steady
state system under various developments of the infall rate
and the modifications to the results if a steady state is
maintained.

We will consider two possible scenarios in which we begin with
a steady state system and then modify the infall rate:  a very high
initial rate ($\dot{M} \gtrsim 10^3 \, M_{\odot} {\rm y}^{-1}$) corresponding
to supernova fallback, and a low initial rate ($\dot{M} \lesssim 10^{-4}
\, M_{\odot} {\rm y}^{-1}$) corresponding to stellar encounters.
For the very high initial rate, a steady state system can not be formed
because the convection timescale is longer than the timescale for the
neutron star to accrete sufficient material to become a black hole.
For low initial infall rates, a steady state system can form and be
maintained with a sufficiently slow increase in the infall rate.

As we increase the infall rate on an initial equilibrium atmosphere,
the entropy of the material at the shock radius decreases, becoming
lower than the entropy of the equilibrium atmosphere.
This system is then unstable to convection.  If convection can equilibrate
this disparity in entropy before the entropy changes sufficiently to
cause vigorous convection, then the system will remain in steady state.
We can estimate a minimum timescale required for the convection from
our calculation of the convective turnover timescale [see eq.
(\ref{eq:tur})].  This timescale was derived assuming the same
vigorous convection which we are trying to avoid and is, therefore,
certainly an underestimate of the time required.  At a radius of
$10^{11} cm$, the vigorous convective timescale is $1.6 \times 10^4 s$.
We will discuss the details of convection in the appendix from
which we determine that for density enhancements greater than $\sim 20\%$,
the convective velocity rises within a factor of 2 of the sound speed,
which is too fast to model under the mixing length algorithm.  We will
define ``vigorous'' convection to begin where mixing length fails.
With this approximation and using equation (\ref{eq:rff}),
we notice that only for situations where $\dot{M}$ changes by
less than $20\%$ over the convective timescale can a steady-state
system be maintained.  From Table 1, we see that even this underestimate
of the timescale precludes most collision formation scenarios, but allows
for the possibility to maintain a steady state in common envelope systems.

As mentioned in section \ref{sec:summ}, for $50 < S < 30$
atmospheres, neutrino heating leads to an explosion.  Assuming
that the initial transient defines the entropy profile, these
entropies are achieved for $\dot{M} \gtrsim 0.1\, M_{\odot}
{\rm y}^{-1}$.  In the steady state solution, we can use
equation (\ref{eq:sin}) and equations (\ref{eq:rsh1}-\ref{eq:rsh3})
to determine the infall rate above which the equilibrium atmosphere
entropy is less than $50$:  $\dot{M_{crit}} = 20 \, M_{\odot}
{\rm y}^{-1}$.  Thus, if a steady state is maintained, the critical
infall rate between the low (hypercritical accretion) regime
and the medium-low (explosion) regime will move from $0.1 \, M_{\odot}
{\rm y}^{-1}$ to $20 \, M_{\odot} {\rm y}^{-1}$.

\section{IMPLICATIONS}

We can now apply the results from the two sets of simulations to a
range of astrophysical situations involving high infall rates onto
neutron stars. In this section, we discuss neutron star accretion in
the context of TZ objects, common envelope systems, neutron stars
in dense molecular clouds, and supernovae.  Because our two-dimensional
simulations invalidate a number of results previously obtained in one
dimension, we present a criterion for the appropriateness of the mixing
length algorithm to model convection. Further, the discussion of specific
systems provides examples of the methodology in applying our results to
the study of other objects. We end this work with a brief discussion
of the observational properties of the explosion regime and a note
on plans for future work.

\subsection{TZ Objects}

All the formation scenarios for TZ objects discussed previously
(see \S \ref{sec:int}) involve a neutron star spiralling into
a red giant star.  As it approaches the core, the infall rate
becomes high.  For example, a neutron star moving at 100 km s$^{-1}$
at a radius of $5 \times 10^{10}$ cm inside a 20 $M_{\odot}$ giant
has a mass infall rate of $1.8\times10^{6}\, M_\odot$ y$^{-1}$.
We can easily verify that all our assumptions hold. That is,
the time delay due to angular momentum transport is short
($\sim 0.5$ s) and the impact of rotational support is minimal
(see \S \ref{sec:ang}), magnetic fields $\lesssim 10^{15}$ Gauss
will be buried by the inflow of material (see \S \ref{sec:mag}),
and photons will be trapped out to the Bondi accretion radius
(see \S \ref{sec:pho}).  Looking at Tables 3 and 4, we see that
this infall fits into the intermediate regime, which forms
a shocked atmosphere but accretes it through neutrino emission
before it can become completely mixed.  These kind of systems
quickly collapse into black holes (for our specific case, the time
scale for collapse $\sim 1$ minute).  Hence, the current range
of scenarios cited in the literature as possible birth mechanisms
for TZ objects will {\it not} form TZ objects.

Even assuming that a proper formation scenario can be found,
it is difficult to imagine how a TZ object could exist for an
extended period of time. Peculiarities in the structure of TZO
seem to inevitably lead to instabilities which destroy the object.
All stellar structure models of TZO have to smoothly connect
the base of the envelope to the surface of the neutron star.
In order to prevent significant neutrino emission, the base of
the atmosphere must remain relatively cool ($\lesssim 10^9 K$).
Eich et al. (1989) were able to construct such cool inner regions
(which they call insulating layers) with low neutrino emission
while maintaining the appropriate pressures.  They argued that
high-temperature atmospheres would emit neutrinos and turn
into the low-temperature stable atmospheres that they had
created.  Our results clearly show that this is not the case
and that once it begins, neutrino emission increases to a
high value which maintains a high rate of accretion.
In our simulations, after an initial transient, the neutrino
emission rate becomes nearly constant (see Fig. 13) rather
than shutting itself off after cooling the material at the
surface of the neutron star.

This can readily be explained by the fact that energy
losses due to neutrinos deleptonize and decrease the specific
internal energy of the base of the atmosphere. Since the
pressure is set by the structure above, the base of the
atmosphere can only adjust and try to maintain pressure
equilibrium by compression. This increases density
{\it and temperature} (or if degeneracy has set in, the
Fermi energy increases) and thus keeps up the neutrino
emission rate until the material is incorporated into the
neutron star. It may appear paradoxical that a loss of
energy via neutrino emission could increase the temperature
(or the Fermi energy).  However, we all know that as a
star evolves, the entropy of the core continually decreases,
while the central temperature keeps increasing [see also eq.
(\ref{eq:tatm})]. All this is related to the fact that under
certain conditions, the heat capacity of gravitational systems
can be negative (inasmuch as a gravitational system can be
considered a thermodynamical system). As a result, once
neutrino emission begins to have a dynamical effect, i.e.
the compression of the base of the atmosphere, it will
continue to be important.  Thus the low temperature
atmospheres constructed by Eich et al. (1989) are unstable.

In addition, the region above the inner layer postulated by Eich
et al. (1989), has to have a high entropy $S > 600$ $k_B$/nucleon,
so that neutrino losses remain unimportant (Fig. 14). Not withstanding
the fact that such entropies are an order of magnitude greater
than those found in main sequence or even giant stars, a
gravitationally bound atmosphere with $S>600$ would have
little mass (recall Fig. 8).  Biehle (1991, 1994), Cannon et
al. (1992), and Cannon (1993) have attempted to overcome this mass
problem by placing a low-entropy envelope on top of the
high-entropy inner region, with associated large negative
entropy gradients (see Fig. 16).  These atmospheres are, of
course, unstable and Biehle and Cannon use the convective
instability to transport energy outward and bring fuel down
into the burning region of their stellar models.  However,
they treat convection with the mixing length approximation.
While Biehle and Cannon were able to maintain the structure
of their atmospheres by assuming mixing length convection,
we have found that this assumption is invalid by running
two-dimensional calculations.  Figure 17 illustrates the
vigorous convection arising from Biehle's initial structure
after 0.1 s.  This convection eventually drives a shock through
the atmosphere, disrupting it, and blowing it away.

An intrinsic assumption of mixing length theory is that the convective
evolution is nearly adiabatic, and slow compared to the dynamical timescale
so that the evolution can be represented by a series of
quasi-static equilibria. As a result, a necessary (but probably
not sufficient) criterion for the validity of mixing length
is that the sound travel time across a convective cell is
much less than the rise time for that cell to move one cell
length, or equivalently that the convective velocity is very subsonic.
The mathematical details of this criterion are discussed in the
appendix.  Table 6 lists the rise times and velocities calculated with
equations (\ref{eq:vcell}) and (\ref{eq:xcell}) for the convective cells after
they travel one scale height for a typical supernovae
simulation, the Sun, and the TZ models of Biehle and Cannon.
Note that only for the Sun is the ratio of the rise time over
the sound crossing time {\it much} less than one.  Of our
four examples, only the Sun satisfies this essential assumption
of mixing length theory.

We have argued above that currently envisioned astrophysical
scenarios are incapable of forming TZ objects. It also appears
that present models of TZ structures improperly account for
convection using mixing-length algorithm, and thus result in
unphysical objects which are artificially stable.

\subsection{Common Envelope Systems}

For a common envelope system, comparison to our results is
less straightforward.  Let us again discuss the characteristics
of a specific case from Table 1.  For a neutron star $10^{12}$ cm
from the center of a $20 M_{\odot}$ giant, moving at
$v \approx v_{\rm orb} = 100$ km s$^{-1}$, the infall rate is $175\,
M_{\odot}$ y$^{-1}$. Again, photons are trapped out to the
Bondi radius and magnetic fields $\lesssim 4 \times 10^{13}$
Gauss will be smothered by the infalling material.  Angular
momentum induces a significant delay time in the accretion
of order 200 s which is nevertheless much less than the
orbital timescale. As we shall see, this situation leads
to a neutrino induced explosion.

The convection timescale for this particular case is much
less than the neutrino cooling timescale (see Table 4).  This
corresponds to the medium-low infall rate regime from our
results.  In this regime, the postshock entropy of the infalling
material is critical to determine the outcome. From
Figure 2, it is clear the accretion shock will lie
at a radius $> 10^8$ cm, and using equation (\ref{eq:sin})
the postshock entropy will be $\lesssim 50$ $k_B$/nucleon,
if the shock is strong, which it is not since this value is
close to the specific entropy of the material outside the
Bondi radius.  Nevertheless, this allows to estimate that
the entropy of the equilibrium atmosphere which forms
above the neutron star corresponds to  $29<S<50$ where
the upper limit comes from the maximum post accretion
shock entropy and the minimum comes from the entropy
of the ambient matter.

{}From our models of equilibrium atmospheres (see \S
\ref{sec:eqatm}), we know that these conditions will lead to
explosions.  These explosions may be sufficient to blow off
the atmosphere and halt the inward spiral of the neutron star,
forming close binary systems such as PSR 1913+16 (see Smarr
\& Blandford 1976 or Burrows \& Woosley 1986).  However, if
the injected energy is insufficient to completely expel the
atmosphere, the neutron star continues to fall into the giant
star, as a new atmosphere once again builds up around it.
Extrapolating from Table 4, we estimate that the neutron star
might survive 50 outbursts over 100 years before it accretes
$\sim 0.2 \; M_{\odot}$.  Simulations of double core evolution
(Terman, Taam, \& Hernquist 1994) estimate inspiral
times $\lesssim 1$ year with some cases where the energy
input from viscous forces on the neutron star ($\sim 10^{47}$
ergs) is sufficient to drive off the envelope and halt
the inward spiral of the neutron star.  The neutron star
will certainly survive this evolution, and this offers yet
another way to form close binary systems.

However, we must here qualify our claims. As we have said before,
it is not possible to run a continuous simulation from the convective
infall regime to the stable equilibrium atmosphere which eventually
appears. Thus, we are forced to infer indirectly the entropy of this
equilibrium atmosphere from the early behavior observed in our
infall simulations. Should the entropy for some reason end up
larger than 50, then an explosion will not occur, but rather, the
atmosphere will undergo steady, hypercritical accretion
until the neutron star collapses to form a black hole.

\subsection{Supernovae}

At present, the details surrounding the explosion mechanism for
supernovae are not sufficiently well understood to place any firm
constraints on the fallback of matter onto the neutron star after
a successful explosion (e.g. see Herant et al. 1994). Taking into
account these uncertainties we would still like to address the
questions whether fallback can lead to the formation of a black
hole or a secondary explosion. A $25 \; M_{\odot}$ supernova
progenitor exploded by Woosley \& Weaver (1995) gives rise to an
initial fallback rate of $10^7\, M_{\odot}$ y$^{-1}$, decreasing
thereafter.  This initial value is just within the high accretion
regime which corresponds to unrestricted accretion by the neutron
star.  This may or may not push the neutron star over its maximum
mass, and make it collapse into a black hole. If this does not
happen, the declining accretion rate will eventually reach the low
infall regime which corresponds to explosions.  Those would then
blow off the remaining bound atmosphere.  Note that the infall rate
from this particular simulation was near the division between the
high and low infall regimes, implying that uncertainties in the
explosion mechanism coupled with differences between supernova
progenitors may lead to very different outcomes, one in which
a black hole forms, and another in which a secondary explosions
expels the remaining material bound to the neutron star.

\subsection{Explosions}\label{sec:exp}

Explosions add an entirely new observational dimension to the
the evolution of rapid mass infall systems which we have
considered in this paper. Physically, these explosions are
most akin to Type II supernovae, and thus, some of the observational
aspects may be similar (velocities, compositions) even though
the amount of mass expelled and energy should be a factor of
$10^{-5}$ smaller. Moreover, it is clear that the extent
and amount of material in which a neutron star is embedded
during such an explosion will have a crucial impact on the
observational signature.  The range of possible signatures is
vast, requiring a more detailed analysis which we relegate
to future work.  However, in some circumstances, these
supernova-like objects may still be bright enough to be seen in
nearby galaxies.

Using observed abundances of Be systems and Massive Binary
systems, Biehle (1991) has derived the formation rate of
common envelope systems to be between $2 \times 10^{-5}$ and
$6 \times 10^{-4}$ per year in our galaxy.  Using the entire
set of observations of massive X-ray binaries, Cannon (1993)
gave a not too different estimate of $10^{-3}$ objects per
year in our galaxy.  Iben, Tututov, \& Yungelson (1995)
predict $1.5 \times 10^{-3}$ objects using an entire
neutron star census.  Because the ``embedding''
companion star will usually be massive, explosions in
these systems are likely to be damped as they propagate
through the massive envelope.  They may therefore appear
only as enhancements of an already strong wind, and changes
in chemical abundances.

Focusing on globular clusters, Davies \& Benz (1995), have
obtained a reliable formation rate of $10^{-8}$ per year
per cluster (which corresponds to $\sim 10^{-6}$ per year per
galaxy) through extensive encounter simulations. They
also predict that these collisions generally result in a
$\sim 0.3\, M_{\odot}$ atmosphere remaining bound to the neutron
star.  Due to the smaller amount of mass surrounding the neutron
star, explosions from these atmospheres will be less damped
and might be observed as supernova-like objects.  However,
the low formation rates limit the observational prospects
for these objects.  One should note though, that these objects
have been proposed as progenitors of millisecond pulsars.
If, instead, they blow off their atmospheres before accreting
sufficiently to be spun up, other scenarios for the production
of millisecond pulsars will have to be found.  Further study
including angular momentum effects will better address this
problem.  Similarly to globular clusters, as dense stellar
systems, galactic bulges offer opportunities for mergers
involving neutron stars through collisions.  However,
we are not aware of reliable estimates for collisional
rates in the galactic center.

Leonard et al. (1994) investigated the scenario in which the
velocity kick received by a neutron star in a supernova
explosion makes it merge with a binary companion. They predicted
an occurence rate of $2.5 \times 10^{-4}$ per year in our galaxy.
These systems may result in an inward spiralling neutron star.
However, since the kick produces collisions with similar
velocities to those from the globular cluster collisions
of Davies \& Benz, we might instead expect the likely result to be
$\sim 0.3\, M_{\odot}$ smothered neutron stars as predicted in
their models.  Assuming the latter to be the case, we would
expect $1\%$ of observed supernovae to produce a secondary explosion
and lead to a peculiar structure of the remnant.

\subsection{Future Work}

We would like to follow-up the discovery of these ``accretion
induced'' explosions with detailed calculations of their observable
signatures.  The observational prospects for the ejection of neutron star
atmospheres are tightly linked to the fraction of encounters
which result in energetic explosions.  This fraction is in
turn strongly dependent on the effects of angular momentum.
While at best difficult, accounting for those effects will
be essential to understanding neutron stars accreting at high
rates.  We would also like to determine the observational
properties and chemical composition of the ejected material.
This will be addressed in future work.

Although it is an interesting topic, we have not discussed the
case of neutron stars embedded in dense molecular clouds or in
AGN disks elsewhere in this paper. Unfortunately, in these
conditions, the photon trapping radius is within the Bondi radius
so that radiation transport plays an important role in the evolution.
It is plausible that dense molecular cloud will first accrete
slowly, at the Eddington rate, until sufficiently high pressures
and temperatures near the neutron star surface develop, leading
then to hypercritical accretion or possibly an expulsion of
the material.  This ejecta will enrich the surrounding medium
and, since the mass accretion will be low, may be a repeatable
process, facilitating an important mechanism to enrich the
interstellar medium or the disk of an AGN. Understanding these
effects will require the implementation of a radiation transport
scheme.

\acknowledgements
This paper has benefited from the contributions of many people.
We are grateful to Raph Hix for making available his nuclear
statistical equilibrium code, to Doug Swesty for his nuclear
equation of state, to Dimitrij Nadezhin for his equation of state,
and to Chuck Wingate for his graphics software. We thank Dave
Arnett for use of his stellar structure code and many helpful
discussions, Stirling Colgate for his advice and for bringing up
many of the questions motivating this paper, and Stan Woosley
for his supernova fallback scenario and related discussions.
Discussions with Grant Bazan, Adam Burrows, Roger Chevalier,
Melvyn Davies and Doug Lin have also served to clarify aspects
of this paper.  The work of C.F. and W.B. was partially supported
by NSF grant AST 9206738 and a ``Profil 2'' grant from the Swiss
National Science Foundation.  The work of M. H. was supported by a Compton
Gamma Ray Observatory postdoctoral fellowship at the University
of California, Santa Cruz, and by a director's postdoctoral
fellowship at the Los Alamos National Laboratory.

\appendix
\section{A Necessary Criterion for the Validity of the Mixing
Length Approximation for Convection}

Despite the many problems with mixing length (choice of scale
height, etc.), there is no better convection algorithm
short of multidimensional simulations.  Hence, mixing length
theory remains the most common technique for dealing with
convective instabilities.  It is thus worthwhile to try to
derive a simple criterion to verify the validity of a mixing
length approach in a given situation.  An intrinsic assumption
of mixing length theory is that the convective evolution is
nearly adiabatic, and slow compared to the dynamical timescale
so that the evolution can be represented by a series of
quasi-static equilibria. As a result, a necessary (but probably
not sufficient) criterion for the validity of mixing length
is that the sound travel time across a convective cell is
{\it much} less than the rise time for that cell to move one cell
length, or equivalently that the convective velocity is very subsonic.
In the following paragraphs, we first provide a rigorous calculation
of the motion of a convective cell which we then complement with
a more physically intuitive interpretation. We go on to
show that typical convective neutron star atmospheres do not
satisfy our criterion and therefore cannot be modeled using
mixing length.

The sound travel time across a cell is simply given by:
\begin{equation}
\tau_s = H_p/c_s
\end{equation}
where $H_p$ is the convective scale length, typically approximated
as the pressure scale height and $c_s$ is the sound speed of the
convective cell.  The acceleration for the
cell is [see Hansen \& Kawaler (1994) for a basic summary]:
\begin{equation}
a_{\rm cell} = a_{\rm buo} + a_{\rm vis} = g(1-\rho/\rho_{c}) -
\frac{\eta}{\rho} \nabla^2 v - \frac{\theta_{\rm turb}}{V\rho} \label{eq:bacc}
\end{equation}
where $g$ is the gravitational acceleration at the position of the
cell, $\rho_{c}$ is the density of the cell, $\rho$ is the
density of the medium through which the cell is passing and $V $
is the volume of the cell.  The quantity $\eta$ is the linear
viscous term and can be written (see Kippenhahn \& Weigert 1990):
\begin{equation}
\eta \approx \eta_{\rm th} + \eta_{\rm rad} \approx
\rho l_{\rm mfp} v_{\rm th} + a T^4/ c \kappa \rho
\end{equation}
where $l_{\rm mfp}$ is the electron mean free path and $v_{\rm th}$ is
the electron's thermal velocity.  We shall approximate
$\nabla^2 v = v/(f H_p)^2$ where $f < 1$ (in our calculations,
we choose $f=0.01$).  The term $\theta_{\rm turb}$ is
the turbulent drag [see, for example, Shames (1992)]:
\begin{equation}
\theta_{\rm turb} \approx \frac{C_{\rm turb} \rho v^2 A}{2} \label{eq:turb}
\end{equation}
where $C_{\rm turb}$ depends upon how streamlined our convective
cells are (values range from 0.01--1 and we conservatively use 1)
and $A$ is the effective surface area of the cell.

We can then integrate this equation to determine the velocity of
the cell (assuming it starts at rest) and distance as a function
of time:
\begin{equation}
v = (L+M) \left( \frac{e^{2 L c t} - 1}{\left( \frac{L+M}{L-M}
\right)e^{2 L c t} + 1} \right), \label{eq:vcell}
\end{equation}
\begin{equation}
x = -\ln \left( \frac{ \left( \frac{L+M}{L-M} \right) e^{2 L c t}
+ 1}{c} \right) - (L+M) t + \ln \left( \frac{ \left( \frac{L+M}{L-M}
\right) + 1}{c} \right) \label{eq:xcell}
\end{equation}
where $L=\sqrt{b^2/(4 c^2) - a/c}$ and $M=b/2c$
with $a=a_{\rm buo}$, $b=-\eta/(f \rho H_p^2)$, and
$c=-(C_{\rm turb} \rho A)/(2 \rho_c)$.  Setting $x=H_p$, one
can solve for the rising time scale, and determine whether
the evolution can be appropriately modeled with a mixing length
algorithm.

In most circumstances, the dominant viscous force arises from the
turbulent drag term.  Ignoring the linear viscous terms simplifies
the preceding equations and provides a more intuitive picture of
the conditions required for using the mixing length formalism
(although we recommend the general argument for any applications).
As we stated earlier, a necessary condition for mixing length
is that the sound travel time is much less than the convective
travel time, or equivalently, the convective velocity must be
much slower than the sound speed.  By eliminating the linear term
from equation (\ref{eq:bacc}), using equation (\ref{eq:turb}) with $A/V
= 1/H_p$, and setting $g = \nabla P/\rho = P/(\rho H_p)$, we can
solve for the maximum bubble velocity:
\begin{equation}
v_{\rm max}=\sqrt{\frac{2 P}{\rho C_{\rm turb}} (\rho/\rho_c
 - 1)} =c_s \sqrt{\frac{2}{C_{\rm turb} \gamma}
(\rho/\rho_{c} - 1)}.
\end{equation}
For a $\gamma = 4/3$ gas and $\rho_{c} < .85 \rho$, we find
that $v_{\rm max} > .5 c_s$, violating the mixing length assumption
that $v \ll c_s$.  However, these high velocities are only a
problem if the bubble can attain them before dispersing.
We can estimate the velocity of the bubble after travelling
a distance $d$ by assuming that the turbulent viscosity is small
until $v$ approaches $v_{\rm max}$:
\begin{equation}
v = \sqrt{2 d a} = c_s \sqrt{\frac{d}{H_p} \frac{2}
{C_{\rm turb}\gamma} (\rho/\rho_{\rm c} - 1)}.
\end{equation}
Setting $d = H_p$ gives $v \rightarrow v_{\rm max}$.
It is likely, then, that the bubble will approach its
maximum velocity after rising one scale length.
The ratio of the bubble density to the density of the
ambiant medium is clearly the primary parameter
behind this necessary criterion for mixing length and
can be simply applied to any system.  One merely needs
to determine the buoyancy (or density) of a bubble
raised adiabatically one scale height.

\clearpage

{}~~~~~~~~~~~~~~~~~~~~~~~~~~~~~~~~~~~~~~~~~~~{\bf Figure Captions}

Figure 1 - Characteristic timescales; photon diffusion $\tau_{\gamma}$,
neutrino cooling $\tau_{\nu}$, and sound travel time $\tau_{s}$
versus entropy.  Note that at an entropy lower than that defined
by the intersection between the sound travel and neutrino
cooling timescale, no stable atmosphere can form. Note also that
the intersection between photon diffusion and neutrino cooling
defines the entropy at which photon trapping is complete.

\bigskip

Figure 2 - Structure of an infall atmosphere.  Note that the region
within the accretion shock is convectively unstable.

\bigskip

Figure 3 - Density vs. radius for a $10^3 M_{\odot}$ y$^{-1}$
infall model.  The points are from numerical simulations.  The
lines are analytical results for different adiabatic indices.

\bigskip

Figure 4 - Steady state shock radius versus accretion
rate for a range of adiabatic indices in a one-dimensional
infall model.

\bigskip

Figure 5 - Entropy versus radius after 50 ms for a range of infall
atmospheres.

\bigskip

Figure 6 - Entropy-driven convective plume for $10^3 M_{\odot}$
y$^{-1}$ infall atmosphere 10 s after creation of the accretion
shock.  The negative entropy gradient is induced by the initial
outward motion of the shock.  Average plume velocity is 3000 km/s
and mean inflow velocity is 1000 km/s.

\bigskip

Figure 7 - Atmosphere mass versus entropy for two sizes of atmosphere:
$R=10^9$ cm and $10^{13}$ cm.  The dashed lines denote atmospheres with
no external pressure, whereas the solid lines include a pressure term
derived from typical values for Bondi-Hoyle accretion.

\bigskip

Figure 8 - Entropy profiles at 70 ms intervals for an $S_{\rm tot} = 50$
equilibrium atmosphere.  Note that with increasing time, the innermost
material cools (lowering its entropy) while an increasing amount of material
is heated (raising its entropy).

\bigskip

Figure 9 - Entropy-driven convection for an $S_{\rm tot} = 50$
equilibrium atmosphere 200 ms after turning on neutrino processes.
The negative entropy gradient is induced by neutrino heating and
drives convection.  The mean outflow velocity is 4000 km/s and the
mean inflow is 9000 km/s.

\bigskip

Figure 10 - The same simulation from Figure 4 after 500 ms.  The
atmosphere is now exploding with a kinetic energy of $2 \times
10^{-6} $ foe.  The mean outflow velocity is 5000 km/s and
the mean inflow is 3000 km/s.

\bigskip

Figure 11 - Entropy vs. mass at 80 ms intervals for an $S_{\rm tot} = 80$
equilibrium atmosphere.  Note that the inner material radiates its energy
through neutrino losses, lowering its entropy.

\bigskip

Figure 12 - Neutrino luminosity and mean neutrino energy
as a function of time for an $S_{\rm tot} = 80$ equilibrium atmosphere.

\bigskip

Figure 13 - Total mass accreted versus time with the best constant
accretion fit for an $S_{\rm tot} = 80$ equilibrium atmosphere.

\bigskip

Figure 14 - A comparison between numerical accretion rates and the
analytical results of Colgate et al. (1993).  The filled circles
compare analytical results using only the radiation entropy from the
numerical calculations, whereas the open symbols use the total entropy.

\bigskip

Figure 15 - Accretion rate versus entropy.  The two lines denote
analytical results using outer pressures of 0\% and 0.25\% that of
the pressure at the surface of the neutron star.  The points are
results from the simulations.

\bigskip

Figure 16 - Entropy profiles versus radius for the nucleosynthesis models
of Cannon (top line) and Biehle (bottom line).

\bigskip

Figure 17 - Biehle's structure model after 0.5 s.  The high entropy
material is driving the atmosphere outward.  Outward velocities
approach 6000 km/s.


\begin{thebibliography}{}

\bibitem[Benz 1991]{Ben91}
Benz, W. 1991, An introduction to Computational Methods in
Hydrodynamics, in Late Stages of Stellar Evolution and Computational
Methods in Astrophysical Hydrodynamics, ed. C.B. de Loore,
(Berlin:Springer), p 259

\bibitem[Benz and Hills 1992]{Ben92}
Benz, W., \& Hills, J.G. 1992, ApJ, 389, 546

\bibitem[Bethe 1990]{Bet90}
Bethe, H.A. 1990, Rev. of Mod. Phys., 62, 801

\bibitem[1991]{Bie91}
Biehle, G.T. 1991, ApJ, 380, 167

\bibitem[1994]{Bie94}
Biehle, G.T. 1994, ApJ, 420, 364

\bibitem[1984]{Bis84}
Bisnovatyi-Kogan, G.S., \& Lamzin, S.A. 1984, Soviet Astron., 28, 187

\bibitem[1995]{Bli95}
Blinnikov, S.I., Dunina-Barkovskaya N.V., \& Nadezhin D.K. 1995,
A\&A, submitted

\bibitem[Bondi 1952]{Bon52}
Bondi, H. 1952, MNRAS, 112, 195

\bibitem[Brown 1995]{Bro95}
Brown, G.E. 1995, ApJ, 440, 270

\bibitem[1986]{Bur86}
Burrows, A., \& Woosley, S. 1986, ApJ, 308, 680

\bibitem[1993]{Can93}
Cannon, R.C. 1993, MNRAS, 263, 817

\bibitem[1992]{Can92}
Cannon, R.C., Eggleton, P.P., Zytkow, A.N., \& Podsiadlowski, P. 1992,
ApJ, 386, 206

\bibitem[1989]{Che89}
Chevalier, R.A. 1989, ApJ, 346, 847 (C89)

\bibitem[1993]{Che93}
Chevalier, R.A. 1993, ApJ, 411,L33

\bibitem[Colgate 1971]{Col71}
Colgate, S.A. 1971, ApJ, 163, 221

\bibitem[1993]{Col93}
Colgate, S.A., Herant, M., \& Benz, W. 1993, Phys. Rep., 227, 157 (CHB)

\bibitem[1983]{Cox83}
Cox, A.N., Vauclair, S., \& Zahn, J.P. 1983, Astrophysical Processes in
Upper Main Sequence Stars, (CH-1290 Sauverny : Geneva Observatory)

\bibitem[1980]{Dav80}
Davies, R.E., \& Pringle, J. 1980, MNRAS,  191, 599

\bibitem[1995]{Dav95}
Davies, M.B. \& Benz, W., 1995, MNRAS, submitted

\bibitem[Gamow 1937]{Gam37}
Gamow, G. 1937, Structure of Atomic Nuclei and Nuclear Transformations,
Oxford University Press

\bibitem[1989]{Eich}
Eich, C., Zimmerman, M.E., Thorne, K.S., \& Zytkow, A.N. 1989,
ApJ, 346, 277

\bibitem[Frail, Gross, \& Whiteoak 1994]{Fra94}
Frail, D.A., Gross, W.M., \& Whiteoak J.B.Z. 1994, ApJ,

\bibitem[1988]{Fry88}
Fryxell, B.A., \& Taam, R.E. 1988, ApJ, 335, 862

\bibitem[1994]{Han94}
Hansen, C.J., Kawaler, S.D., Stellar Interiors-Physical Principles,
Structure, and Evolution, 1994, (New York: Springer-Verlag)

\bibitem[1994]{Her94}
Herant, M., Benz, W., Hix, W. R., Fryer, C. L., \& Colgate, S. A.
1994, ApJ, 435, 339 (HBHFC)

\bibitem[1992]{Her92a}
Herant, M. \& Benz, W. 1992, ApJ, 387, 294

\bibitem[1992]{Her92b}
Herant, M., Benz, W., \& Colgate, S.A. 1992, ApJ, 395, 642

\bibitem[1995]{Hix95}
Hix, R., Thielemann, F.K., Fushiki, I., \& Truran, J.W. 1994,
in preparation for ApJ

\bibitem[1995]{Ibe95}
Iben, I., Tutukov, A.V., Yungelson, L.R., 1995, submitted to ApJ

\bibitem[1991]{Hou91}
Houck, J.C., \& Chevalier, R.A. 1991, ApJ, 376, 234 (HC)

\bibitem[1990]{Kip90}
Kippenhahn, R., \& Weigert, A., 1990, Stellar Structure and Evolution,
(Berlin: Springer-Verlag)

\bibitem[Landau 1937]{Lan37}
Landau, L.D. 1937, Doklady Akad. Nauk USSR, 17, 301

\bibitem[1991]{Lat91}
Lattimer, J. M., Swesty, F. D. 1991, Nuc. Phys. A, 535, 331

\bibitem[Leonard et al. 1994]{Leo94}
Leonard, P.J., Hills, J.G., \& Dewey, R.J. 1994, ApJ, 423, L19

\bibitem[Lyne \& Lorimer 1994]{Lyn94}
Lyne, A.G., \&  Lorimer, D. R. 1994, Nature, 369, 127

\bibitem[1994]{Nad94}
Nadezhin, D.K. 1974,
Naucnye Informatsii Astron. Sov. Akad. Nauk SSR, 32, 33

\bibitem[1994a]{Ruf1}
Ruffert, M. 1994a, ApJ, 427, 342

\bibitem[1994b]{Ruf3}
Ruffert, M. 1994b, A\&AS, 106, 505

\bibitem[1995]{Ruf4}
Ruffert, M. 1995, submitted to ApJ

\bibitem[1995]{Ruf5}
Ruffert, M., \& Anzer, U. 1995, A\&A, 295, 108

\bibitem[1994]{Ruf2}
Ruffert, M., \& Arnett, W.D. 1994, ApJ, 427, 351

\bibitem[1989]{Saw89}
Sawada, K., Matsuda, T., Anzer, U., B\"orner, \& Livio, M.
1989, ApJ, 221, 263

\bibitem[Schinder 1990]{Sch90}
Schinder, P.J. 1990, ApJ, 374, 249

\bibitem[1976]{Sma76}
Smarr, L. L., \& Blandford, R. 1976, ApJ, 207, 574

\bibitem[1992]{Sha92}
Shames, I. H., 1992, Mechanics of Fluids, (New York: McGraw-Hill)

\bibitem[1983]{Sha83}
Shapiro, S.L., \& Teukolsky, S.A., 1983, Black Holes, White Dwarfs,
and Neutron Stars, (New York: Wiley), 405

\bibitem[1978]{Taa78}
Taam, R.E., Bodenheimer, P.,\& Ostriker, J.P. 1978, ApJ, 222, 269

\bibitem[1988]{Taa89}
Taam, R.E., \& Fryxell, B.A., 1989, ApJ, 339, 297

\bibitem[1994]{Ter94}
Terman, J.L., Taam, R.E.,\& Hernquist, L., 1994, ApJ, 422, 729

\bibitem[1993]{The93}
Theuns, T., \& Jorissen, A. 1993, MNRAS, 265, 946

\bibitem[1975]{Tho75}
Thorne, K.S., \& Zytkow, A.N. 1975, ApJ, 199, L19

\bibitem[Thorne and Zytkow 1977]{Tho77}
Thorne, K.S., \& Zytkow, A.N. 1977, ApJ, 212, 832

\bibitem[Van Riper 1979] {Van79}
Van Riper, K.A. 1979, ApJ, 232, 558

\bibitem[Woosley and Weaver] {Woo83}
Woosley, S.E., \& Weaver, T.A. 1983, AIP Conf. Proc., 115, 273

\bibitem[Woosley and Weaver] {Woo95}
Woosley, S.E., \& Weaver, T.A. 1995, ApJ Sup., in press

\bibitem[Zel'dovich et al. 1972]{Zel72}
Zel'dovich, Ya.B., Ivanova, L.N., \& Nadezhin, D.K. 1972,
Soviet Astr.-AJ, 16, 209

\end{thebibliography}
\end{document}